\documentclass[12pt]{article}
\usepackage[latin1]{inputenc} 
\usepackage{amssymb}
\usepackage{amsfonts} 
\usepackage{graphicx} 
\usepackage{amsthm} 
\usepackage{amsmath} 
\usepackage{epsfig}

\usepackage{hyperref}
\def\ni{\noindent}
\def\nn{\nonumber}

\def \bc {\begin{center}}
\def \ec {\end{center}}
\def \bi {\begin{itemize}}
\def \ei {\end{itemize}}

\def \ba {\begin{array}}
\def \ea {\end{array}}

\def \bea {\begin{eqnarray}}
\def \eea {\end{eqnarray}}

\def \be {\begin{equation}}
\def \ee {\end{equation}}

\def \lb {\left[}
\def \rb {\right]}

\newcommand{\la}{\langle}
\newcommand{\ra}{\rangle}

\def\cD {{\cal D}}

\def\cM {{\cal M}}
\def\cU {{\cal U}}
\def\cA {{\cal A}}
\def\cT {{\cal T}}
\def\cN {{\cal N}}
\def\cS {{\cal S}}
\def\cV {{\cal V}}

\def\mbX {{\mathbb X}}
\def\mbN {{\mathbb N}}
\def\mbD {{\mathbb D}}
\def\mbS {{\mathbb S}}
\def\mbC {{\mathbb C}}

\def\tr {{\rm tr}}
\def\um {\frac{1}{2}}

\newtheorem{thm}{Theorem}[section]

\newtheorem{prop}[thm]{Proposition}
\newtheorem{defn}[thm]{Definition}

\begin{document}
\hfill DOI: 10.1016/j.acha.2010.11.004
\begin{center}
{\Large {\bf Extended MacMahon-Schwinger's Master  Theorem  and
 Conformal Wavelets in  Complex Minkowski Space}}
\end{center}
\bigskip

\centerline{{\sc M. Calixto}$^{1,2}$\footnote{Corresponding author:
calixto@ugr.es, Manuel.Calixto@upct.es} and {\sc E. Pérez-Romero}$^{2}$ }

\bigskip

\bc {\it $^1$ Departamento de Matemática Aplicada,
Facultad de Ciencias, Campus de Fuentenueva,  18071 Granada, Spain}
\\
{\it $^2$ Instituto de Astrof\'\i sica de Andaluc\'\i a
(IAA-CSIC), Apartado Postal 3004, 18080 Granada, Spain}
\ec


\bigskip
\begin{center}
{\bf Abstract}
\end{center}
\small
\begin{list}{}{\setlength{\leftmargin}{3pc}\setlength{\rightmargin}{3pc}}
\item We construct the Continuous Wavelet Transform (CWT) on the
homogeneous space (Cartan domain) $\mathbb
D_4=SO(4,2)/(SO(4)\times SO(2))$ of the conformal group $SO(4,2)$
(locally isomorphic to $SU(2,2)$) in 1+3 dimensions. The manifold
$\mathbb D_4$ can be mapped one-to-one onto the future tube domain
$\mbC^4_+$ of the complex Minkowski space through a Cayley
transformation, where other kind of (electromagnetic) wavelets
have already been proposed in the literature. We study the unitary
irreducible representations of the conformal group on the Hilbert
spaces $L^2_h(\mbD_4,d\nu_\lambda)$ and
$L^2_h(\mbC^4_+,d\tilde\nu_\lambda)$  of square integrable
holomorphic functions with scale dimension $\lambda$ and
continuous mass spectrum, prove the isomorphism (equivariance)
between both Hilbert spaces, admissibility and tight-frame
conditions, provide reconstruction formulas and orthonormal
basis of homogeneous polynomials and discuss symmetry properties and the Euclidean limit of the
proposed conformal wavelets. For that purpose, we firstly
state and prove a $\lambda$-extension of Schwinger's Master
Theorem (SMT), which turns out to be a useful mathematical tool
for us, particularly as a generating function for the
unitary-representation functions of the conformal group and for
the derivation of the reproducing (Bergman) kernel of
$L^2_h(\mbD_4,d\nu_\lambda)$. SMT is related to MacMahon's Master
Theorem (MMT) and an extension of both in terms of Louck's $SU(N)$
solid harmonics is also provided for completeness. Convergence
conditions are also studied.
\end{list}
\normalsize 


%
%
%
%
%


\newpage
\section{Introduction}

Since the pioneer work of Grossmann, Morlet and Paul
\cite{GMP}, several extensions of the standard Continuous Wavelet Transform (CWT) on $\mathbb R$
(traditionally based on the affine group of time translations and
dilations, see e.g. \cite{MallatBook,KaiserBook}) to general
manifolds $\mathbb X$ have been constructed (see e.g.
\cite{Gazeau,Fuhr} for general reviews and
\cite{CWTmanifolds,Fuhr2} for recent papers on WT and Gabor
systems on homogeneous manifolds). Particular interesting examples
are the construction of CWT on spheres $\mathbb{S}^{N-1}$, by
means of an appropriate unitary representation of the Lorentz
group  in $N+1$ dimensions $SO(N,1)$
\cite{Holschneider-sphere,waveS2,nsphere}, and on the upper sheet
$\mathbb H^2_+$  of the two-sheeted hyperboloid $\mathbb H^2$
\cite{cwthyperbol}, or its stereographical projection onto the
open unit disk
\be \mathbb D_1=SO(1,2)/SO(2)=SU(1,1)/U(1).\label{unitdisk}\ee
The basic ingredient in all these constructions is
a group of transformations $G$ which contains dilations and motions on
$\mathbb X$, together with a transitive action of $G$ on $\mathbb X$.

In this article we shall consider the 15-parameter conformal group
$G=SO(4,2)$ in 1+3 dimensions and its natural action on the Minkowski
spacetime. The fact that the conformal group contains space-time dilations
and translations leads to a natural generalization of the standard CWT for
signals on the real line to a higher dimensional manifold. Actually, the
conformal group $SO(4,2)$ consists of Poincaré transformations
(space-time translations and Lorentz relativistic rotations and boosts),
augmented by dilations and relativistic uniform accelerations, which can
also be seen as a sort of local (point-dependent) scale transformations
(see later on Section \ref{CWS}).

The conformal group $SO(4,2)$ (or its four-covering $SU(2,2)$) has
been recognized as a symmetry of Maxwell theory of
electromagnetism without sources since \cite{Cunnigham,Bateman}.
Electromagnetic waves turn out to be written as superpositions of
a particular set of \emph{conformal wavelets}
\cite{Kaiser1,Kaiser2,Kaiser3}. Thus, conformal wavelets provide a
local spacetime-scale analysis of electromagnetic waves in much
the same way as standard wavelets provide a time-scale analysis of
time signals. In these works, electromagnetic waves are
analytically continued or extended from real to complex spacetime
and they are obtained from a single \emph{mother} wavelet by
applying conformal transformations of space and time.

Here we shall deal with a different type of conformal wavelets,
although we shall work in complex spacetime too. Besides the above
massless representations of $SO(4,2)$ on the electromagnetic
field, the conformal group has other representations with
continuous mass spectrum labelled by the representations of the
stability subgroup $SO(4)\times SO(2)$: the two \emph{spins}
$s_1,s_2\in\mathbb N/2$ and the \emph{scale dimension} $\lambda\in
\mathbb N$ of the corresponding field \cite{Ruhl}. We shall restrict
ourselves to scalar fields ($s_1=s_2=0$) for the sake of
simplicity. After a reminder of these representations, we provide
admissibility conditions, tight wavelet frames and reconstruction
formulas for functions on the complex Cartan domain or Lie ball
(see \cite{Coquereaux} for a general discussion on these classical
complex domains)
\be \mbD_4=SO(4,2)/(SO(4)\times SO(2))=SU(2,2)/S(U(2)\times
U(2)),\label{4unitdisk}\ee
which is the four-dimensional analogue of the open unit disk
$\mathbb D_1$ abovementioned. This domain can be mapped one-to-one
onto the forward/future tube domain $\mbC^4_+$ (the
four-dimensional analogue of the Poincaré/Lobachevsky/hyperbolic
upper half-plane $\mbC_+$)  of the complex Minkowski space through
a Cayley transformation. For completeness, we also provide an
isometric (equivariant) map between the Hilbert spaces of holomorphic functions
on $\mbD_4$ and $\mbC^4_+$, where we enjoy more physical
intuition.

In order to prove admissibility conditions and reconstruction
formulas, an extension of the traditional Schwinger's Master
Theorem (SMT) \cite{Schwinger} will show up as a useful
mathematical tool for us. Schwinger's inner product formula turns
out to be essentially equivalent to MacMahon's Master Theorem
(MMT) \cite{MacMahon}, which is one of the fundamental results in
combinatorial analysis. A quantum analogue of the MMT has also
been constructed \cite{qMacMahon} and related to a quantum
generalization of the boson-fermion correspondence of Physics.
Moreover, an extension of the classical MMT \cite{MacMahon} was
proved in \cite{MacMahonext} by using the permutation group. The
unification of SMT and MMT into a single form by using properties
of the so-called $SU(N)$ solid harmonics
\cite{Louck0,Louck1,Louck2} (a generalization of Wigner's
$\cD$-matrices for $SU(2)$, see e.g. \cite{Louck3}), was pointed
out by Louck in \cite{Louck0}. The combined MacMahon-Schwinger's
Master Theorem provides a generating function for the diagonal
elements, the trace, and the representation functions of the
so-called totally symmetric unitary representations of the compact
unitary group $U(N)$ \cite{Louck0,Louck1,Louck2}.

In this article we shall state and prove a $\lambda$-extension of
the SMT by using the abovementioned $SU(N)$ solid harmonics of
\cite{Louck1,Louck2}. This $\lambda$-extension of the SMT will
appear to be useful as a generating function for the
unitary-representation functions of the non-compact special
pseudo-unitary group $SU(N,N)$ and for the computation of the
reproducing (Bergman) kernel. We shall concentrate on the $N=2$
case, i.e., on the conformal group $SU(2,2)$ (the general case
$N\geq 2$ is discussed in the Appendix \ref{sizenap}), which will
be essential in the development of conformal wavelets for fields
with continuum mass spectrum.

The paper is organized as follows. In Section \ref{SM} we remind Schwinger
and MacMahon's Master Theorems and state and prove a $\lambda$-extension
of Schwinger's formula. The generalization to matrices $X$ of size $N\geq
2$ is also discussed for completeness in the Appendix \ref{sizenap}. In
order to be as self-contained as possible, in Section \ref{gbackdrop} we
present the group-theoretical backdrop and leave for the Appendix
\ref{brief} a succinct exposition of the CWT on a general manifold $\mbX$,
collecting the main definitions used in this paper. In Section \ref{afin}
we briefly remind the CWT on $\mathbb R$ and extend it to the Lobachevsky
plane $\mbC_+$ and the open unit disk $\mathbb D_1$. The action of the
affine group on  $\mbC_+$ extends naturally to the conformal group
$SO(1,2)$ of the time axis $\mathbb R$. This will serve us to introduce
and establish a parallelism between standard and conformal wavelets in
complex Minkowski space in Section \ref{CWS}. We shall eventually work in
the Cartan domain $\mathbb D_4$, although we shall provide in Section
\ref{isometrysec} an (intertwiner) isometry  between the Hilbert spaces of
holomorphic functions on $\mathbb D_4$ and the future tube domain
$\mbC^4_+$. The $\lambda$-extended SMT turns out to be a valuable
mathematical tool inside $\mathbb D_4$ for proving admissibility and
tight-frame conditions, reconstruction formulas and reproducing (Bergman)
kernels in Section \ref{WCDsec}. We also discuss symmetry properties and
comment on the Euclidean limit of the proposed wavelets in Sections
\ref{symmetrysec} and \ref{euclideansec}, respectively. Section
\ref{convergence} is devoted to convergence considerations and Section
\ref{conclu} to conclusions and outlook. In Appendix \ref{orthoap} we
prove orthonormality properties of a basis of homogeneous polynomials
introduced in Section \ref{CWS}.

\section{\label{SM}Schwinger's Master Theorem: an Extension}

Schwinger's inner product formula  \cite{Schwinger} can be stated
as follows:

{\thm \label{MSMT2}{\rm (SMT)}. Let $X$ be any $2\times 2$ matrix
$X$ and $Y=tI$, where $t$ is an arbitrary parameter and $I$ stands
for the $2\times 2$ identity matrix. Let us denote by
\bea
\cD^{j}_{q_1,q_2}(X)=\sqrt{\frac{(j+q_1)!(j-q_1)!}{(j+q_2)!(j-q_2)!}}
\sum_{k=\max(0,q_1+q_2)}^{\min(j+q_1,j+q_2)}
\binom{j+q_2}{k}\binom{j-q_2}{k-q_1-q_2}\nn\\ \times  x_{11}^k
x_{12}^{j+q_1-k}x_{21}^{j+q_2-k}x_{22}^{k-q_1-q_2},\label{Wignerf}\eea
the Wigner's $\cD$-matrices for $SU(2)$ \cite{Louck3}, where $j\in
\mbN/2$ (the spin) runs on all non-negative half-integers and
$q_1,q_2=-j,-j+1,\dots,j-1,j$. Then the following identity holds:
\be
\left.e^{(\partial_u:X:\partial_v)}e^{(u:Y^T:v)}\right|_{u=v=0}=
\sum_{j\in\mbN/2}t^{2j}\sum_{q=-j}^j\cD^{j}_{qq}(X)={\det(I-tX)}^{-1}\label{MSMTF}
\ee
where we denote by $(u:X:v)\equiv uXv^T=\sum_{i,j=1}^N u_i x_{ij}
v_j$ and $\partial_{u_i}\equiv \partial/\partial{u_i}$.}

\ni This formula turns out to be essentially equivalent to
MacMahon's Master Theorem:
{\thm\label{MSMT} {\rm (MMT)}  Let $X$ be an $N\times N$ matrix of
indeterminates $x_{ij}$, and $Y$ be the diagonal matrix
$Y\equiv{\rm diag}(y_1,y_2,\dots,y_N)$. Then the coefficient of
$y^\alpha\equiv y_1^{\alpha_1}y_2^{\alpha_2}\dots y_N^{\alpha_N}$
in the expansion of   $\det(I-XY)^{-1}$ equals the coefficient of
$y^\alpha$ in the product
\be \prod_{i=1}^N(x_{i1}y_1+x_{i2}y_2+\dots+x_{iN}y_N)^{\alpha_i}.
\ee}
These abovementioned coefficients can be written in terms of the
so-called $SU(N)$ solid harmonics $\cD^p_{\alpha\beta}(X)$ (see
\cite{Louck1,Louck2} and Appendix \ref{sizenap} for a general
definition). $SU(N)$ solid harmonics (\ref{SolidHarmonics}) are a
natural generalization of the standard Wigner's $\cD$-matrices
(\ref{Wignerf}) to matrices $X$ of size $N\geq 2$. In fact,
replacing $tX$ with $XY$ in (\ref{MSMTF}) and using the
multiplication property
\be \sum_{q'=-j}^j
\cD^{j}_{qq'}(X)\cD^{j}_{q'q''}(Y)=\cD^{j}_{qq''}(XY)\label{Dmultprop}\ee
and the transpositional symmetry
\be \cD^{j}_{qq'}(Y)=\cD^{j}_{q'q}(Y^T),\label{transposprop}\ee
we can restate MMT for $N=2$ as:
\be
\sum_{j\in\mbN/2}\sum_{q,q'=-j}^j\cD^{j}_{qq'}(X)\cD^{j}_{qq'}(Y^T)={\det(I-XY)}^{-1}.\ee
Actually, MMT preceded Schwinger's result by many years. Schwinger
re-discovered the MMT in the context of his generating function
approach to the angular momentum theory of many-particle systems.
The unification into a single form by using properties of the
$SU(N)$ solid harmonics was established by Louck in
\cite{Louck0,Louck1,Louck2}.

Wigner matrices $\cD^j_{qq'}(X)$ are homogeneous polynomials of degree
$2j$ in $x_{kl}$. Inspired by Euler's theorem, we shall define the
following differential operator:
\be D_\lambda f(t)\equiv(\lambda+t \frac{\partial}{\partial
t})f(t),\;\;\lambda\in\mathbb N\ee
which will be useful in the sequel. Now we are in condition to
state and prove an extension of SMT \ref{MSMT2}. For the sake of
completeness, a generalization for matrices $X$ of size $N\geq 2$
is also given in Appendix \ref{sizenap}.

{\thm \label{GMSMT2} {\rm ($\lambda$-Extended SMT)} For every $
\lambda  \in \mathbb{N}, \lambda \geq 2$ and every $2\times 2$
matrix $X$, the following identity holds:
 \bea
 \sum_{j\in\mbN/2}\frac{2j+1}{\lambda-1}\sum^{\infty}_{n=0}t^{2j+2n}
 \binom{n+\lambda-2}{\lambda-2}\binom{n+2j+\lambda -1}{\lambda-2}
 \det(X)^{n}\sum_{q=-j}^j\cD^{j}_{qq}(X)\nn\\ =\det( I-tX)^{-\lambda}.\label{EMSMT}\eea
}
\ni\textbf{Proof:} We start from the basic SMT \ref{MSMT2} and
apply the operator $D_{1}$ on both sides of Eq. (\ref{MSMTF}):
\be
 \sum_{j\in\mbN/2}(2j+1)t^{2j}\sum_{q=-j}^j\cD^{j}_{qq}(X)=\frac{1-t^2\det(X)}{\det(
 I-tX)^{2}}\label{MSMT21}.
\ee
Here we have used that
\be \det(I-tX)=1-\tr(tX)+\det(tX) \label{DTrDet}\nn\ee
and that ${\rm tr}(X)$ and $\det(X)$ are homogeneous polynomials of degree
1 and 2, respectively. Making use of the expansion:
\be\frac{1}{1-t^2\det(X)}=\sum^{\infty}_{n=0}t^{2n}\det(X)^{n},\label{taylorexp}\ee
the expression (\ref{MSMT21}) can be recast as:
 \be
 \sum_{j\in\mbN/2}(2j+1)\sum^{\infty}_{n=0}t^{2j+2n}\det(X)^{n}
\sum_{q=-j}^j\cD^{j}_{qq}(X)=\frac{1}{\det(I-tX)^{2}}.\label{lambda2}
 \ee
This identity is a particular case of (\ref{EMSMT}) for $\lambda=2$. Now
we shall proceed by induction on $\lambda$. Assuming that (\ref{EMSMT}) is
valid for every $\lambda\geq 2$ and applying the operator $D_{\lambda }$
on both sides of the equality (\ref{EMSMT}), we arrive at:
 \bea
 \sum_{j\in\mbN/2}\frac{2j+1}{\lambda-1}\sum^{\infty}_{n=0}
 \binom{n+\lambda-2}{\lambda-2}\binom{n+2j+\lambda -1}{\lambda-2} (\lambda
 +2j+2n)t^{2j+2n}\nn\\ \times
 \det(X)^{n}\sum_{q=-j}^j\cD^{j}_{qq}(X) =\lambda
\frac{1-t^{2}\det(X)}{\det(I-tX)^{\lambda+1}},\nn\eea
where we have made use again of (\ref{DTrDet}). Considering
(\ref{taylorexp}) one more time, we can assemble the previous expression
as:
 \bea
\sum_{j\in\mbN/2}\frac{2j+1}{\lambda-1}\sum^{\infty}_{n,m=0}
 \binom{n+\lambda-2}{\lambda-2}\binom{n+2j+\lambda -1}{\lambda-2} (\lambda +2j+2n)t^{2j+2(n+m)}
 \nn\\ \times \det(X)^{n+m}\sum_{q=-j}^j\cD^{j}_{qq}(X)=
\frac{\lambda}{\det(I-tX)^{\lambda+1}}.\label{f16}\eea
Rearranging series:
\be \sum^{\infty}_{n,m=0}a_n b^{n+m}=
\sum^{\infty}_{n=0}\left(\sum^{n}_{m=0}a_m \right)b^{n},
\label{reseries}\ee
the identity (\ref{f16}) can be recast in the form:
 \bea
 \sum_{j\in\mbN/2}\frac{2j+1}{\lambda-1}\sum^{\infty}_{n=0}\sum^{n}_{m=0}
\binom{m+\lambda-2}{\lambda-2}\binom{m+2j+\lambda-1}{\lambda-2}(\lambda+2j+2m)t^{2j+2n}
\nn\\ \times \det(X)^{n}\sum_{q=-j}^j\cD^{j}_{qq}(X) =\frac{\lambda
}{\det( I-tX)^{\lambda +1}}.\label{lemmap}\eea
It remains to prove the following combinatorial identity:
  \bea
 \frac{1}{\lambda-1}\sum^{n}_{m=0}
\binom{m+\lambda-2}{ \lambda-2 } \binom{m+p+\lambda-1}{ \lambda-2
} (\lambda+p+2m) \nn\\ =\binom{n+\lambda-1}{ \lambda-1 }
\binom{n+p+\lambda}{\lambda-1}. \label{binomrel}\eea
We shall proceed by induction on $n$. Let us define both sides of
the previous equality as the two sequences:
 \bea
 F(n)&=&\frac{1}{\lambda-1}\sum^{n}_{m=0}
\binom{m+\lambda-2}{ \lambda-2}\binom{m+p+\lambda-1}{ \lambda-2}
(\lambda+p+2m),\nn\\ G(n)&=&\binom{n+\lambda-1}{ \lambda-1 }
\binom{n+p+\lambda}{ \lambda-1}. \nn\eea
It is easy to verify that $F(0) = G(0)$.  Assuming that $F(n)= G(n)$, we
ask whether  $F(n+1)=G(n+1)$. Indeed, on the one hand
\bea
 F(n+1)&=&\frac{1}{\lambda-1}\sum^{n+1}_{m=0}
\binom{m+\lambda-2}{ \lambda-2}\binom{m+p+\lambda-1}{
\lambda-2} (\lambda+p+2m)\nn\\
&=&F(n)+\frac{\lambda+p+2(n+1)}{\lambda-1}\binom{n+\lambda-1}{
\lambda-2} \binom{n+p+\lambda}{\lambda-2}\nn\\
&=&F(n)+\frac{(\lambda+p+2n+2)(\lambda-1)^2}{(\lambda-1)(n+1)(n+p+2)}G(n)\nn\\
&=&\frac{(n+1)(n+p+2)+(\lambda+p+2n+2)(\lambda-1)}{(n+1)(n+p+2)}G(n)\nn
 \eea
and on the other hand
 \be
G(n+1)=\binom{n+\lambda}{ \lambda-1}
\binom{n+p+\lambda+1}{\lambda-1}=\frac{(n+p+\lambda+1)(n+\lambda)}{(n+1)(n+p+2)}G(n).\nn
 \ee
Realizing that
\be
(n+1)(n+p+2)+(\lambda+p+2n+2)(\lambda-1)=(n+p+\lambda+1)(n+\lambda)\nn\ee
we arrive at $F(n+1)= G(n+1)$, which proves (\ref{binomrel}). Finally,
inserting (\ref{binomrel}) in (\ref{lemmap}), we conclude that
(\ref{EMSMT}) is valid for $\lambda+1$, thus completing the proof
$\blacksquare$

\section{The Group-Theoretical Backdrop}\label{gbackdrop}

The usual CWT on the real line $\mathbb{R}$ is derived from the
natural unitary representation of the affine or similitude group $G=SIM(1)$ in
the space of finite energy signals $L^2(\mathbb{R}, dx)$ (see
Section \ref{afin} for a reminder). The same scheme applies to the
CWT on a general manifold $\mbX$, subject to the transitive
action, $x\to gx, g\in G, x\in \mbX$, of some group of
transformations $G$ which contains dilations and motions on
$\mbX$. We address the reader to the References \cite{Gazeau,Fuhr} for a nice and thorough exposition
on this subject with multiple examples. For the sake of self-containedness, we also collect in the Appendix \ref{brief} some
basic definitions which are essential for our construction of conformal wavelets.

As already said in the Introduction, the CWT on spheres
$\mbX=\mathbb{S}^{N-1}$ has been constructed in
\cite{Holschneider-sphere,waveS2,nsphere} by means of an appropriate
unitary representation of the Lorentz group  in $N+1$ (space-time)
dimensions $G=SO(N,1)$. The case of $G=SO(2,1)$ is particularly
interesting as it encompasses wavelets on the circle $\mathbb{S}^{1}$ and
on the real line $\mathbb R$, associated to the continuous and discrete
series representations, respectively (see \cite{acha} for a unified
group-theoretical treatment of both type of wavelets inside
$SL(2,\mathbb{R})\simeq SO(1,2)$). The group $SO(1,2)$ (the conformal
group in $0+1$-dimensions) has also been used to construct wavelets on the
upper sheet $\mathbb H^2_+$  of the two-sheeted hyperboloid $\mathbb H^2$
\cite{cwthyperbol}, or its stereographical projection onto the open unit
disk (\ref{unitdisk}).

The (angle-preserving) conformal group in $N$ (space-time) dimensions is
finite-dimensional except for $N=2$. For $N\not=2$, the conformal group
$SO(N,2)$ consists of Poincaré [spacetime translations $b^\mu\in\mathbb
R^N$ and restricted Lorentz $\Lambda^{\mu}_\nu\in SO^+(N-1,1)$]
transformations augmented by dilations ($a\in\mathbb R_+$) and
relativistic uniform accelerations (special conformal transformations
$c^\mu\in\mathbb R^N$) which, in $N$-dimensional Minkowski spacetime, have
the following realization:
\be \ba{ll} x'^\mu =x^\mu+b^\mu, & x'^\mu=\Lambda^{\mu}_\nu(\omega) x^\nu,\\
x'^\mu=ax^\mu,& x'^\mu=\frac{x^\mu+c^\mu x^2}{1+2c x+c^2 x^2},\ea
\label{confact} \ee
respectively. We are using the Minkowski metric
$\eta^{\mu\nu}={\rm diag}(1,-1,\overbrace{\dots}^{N-1},-1)$ to rise and lower
space-time indices and the Einstein summation convention $cx=c_\mu
x^\mu$. The new ingredients with regard to the affine group
$SIM(1)$ are the extension from time-translations by $b^0$ to
$N$-translations by $b^\mu$, the addition of Lorentz
transformations $\Lambda^{\mu}_\nu$ (rotations and boosts) and
accelerations by $c^\mu$. Special conformal transformations can be
seen as a sequence of inversions and translations by $c^\mu$ of
the form:
\be x^{\mu}\stackrel{{\rm
inv}}{\longrightarrow}\frac{x^{\mu}}{x^2}\stackrel{c_\mu}{\longrightarrow}\frac{x^{\mu}+x^2c^{\mu}}{x^2}\stackrel{{\rm
inv}}{\longrightarrow}
\frac{(x^{\mu}+x^2c^{\mu})/x^2}{(x^{\mu}+x^2c^{\mu})^2/x^4}=\frac{x^\mu+c^\mu
x^2}{1+2c x+c^2 x^2}.\label{sctinv}\ee
They can also be interpreted as point-dependent
(generalized/gauge) dilations in the sense that, while standard
dilations change the spacetime interval $ds^2=dx^\mu dx_\mu$
globally as $ds^2\to a^2ds^2$, special conformal transformations
scale the spacetime interval point-to-point as $ds^2\to
\sigma(x)^{-2} ds^2$, with $\sigma(x)=1+2c x+c^2 x^2$. The same
happens with the squared mass $m^2$, thus forcing a continuous
mass spectrum unless $m=0$, as for the electromagnetic field.

The infinitesimal generators of the transformations (\ref{confact}) are
easily deduced:
\be \ba{rcl}P_\mu &=& \frac{\partial}{\partial x^\mu}, \;\;
M_{\mu\nu}=x_\mu
\frac{\partial}{\partial x^\nu}-x_\nu \frac{\partial}{\partial x^\mu},\\
D&=&x^\mu\frac{\partial}{\partial x^\mu},\;\; K_\mu=-2x_\mu x^\nu
\frac{\partial}{\partial x^\nu}+x^2\frac{\partial}{\partial x^\mu},
\label{confvf}\ea\ee
and they close into the conformal Lie algebra
 \be\ba{rcl} \lb M_{\mu\nu},M_{\rho\sigma}\rb &=&\eta_{\nu\rho}M_{\mu\sigma}+\eta_{\mu\sigma}M_{\nu\rho}
-\eta_{\mu\rho}M_{\nu\sigma}-\eta_{\nu\sigma}M_{\mu\rho},\\
\left[P_\mu,M_{\rho\sigma}\right] &=& \eta_{\mu\rho} P_\sigma -
\eta_{\mu\sigma}
P_\rho,\;\; \lb P_\mu,P_\nu\rb=0,\\
\lb K_\mu,M_{\rho\sigma}\rb &=& \eta_{\mu\rho}K_\sigma-\eta_{\mu\sigma}K_\rho,\;\; \lb K_\mu,K_\nu\rb=0, \\
\lb D,P_\mu\rb &=&-P_\mu, \;\;\lb D,K_\mu\rb =K_\mu,\;\; \lb D,M_{\mu\nu}\rb=0,\\
\lb K_\mu,P_\nu\rb &=& 2(\eta_{\mu\nu}
D+M_{\mu\nu}).\ea\label{conformalgebra}\ee
The quadratic Casimir operator:
%
\be C_2=D^2-\um M_{\mu\nu}M^{\mu\nu}+\um(P_\mu K^\mu+K_\mu P^\mu)
,\label{Casimir}\ee
generalizes the Poincaré Casimir $P^2=P_\mu P^\mu$ (the squared rest mass).

Any group element $g\in SO(4,2)$ (near the identity element) could be
written  as the exponential map
\be g=\exp(u),\; u=\tau D+b^\mu P_\mu+c^\mu
K_\mu+\omega^{\mu\nu}M_{\mu\nu},\label{expmap}\ee
of the Lie-algebra element $u$ (see \cite{expomap0,expomap1}). The compactified
Minkowski space  $\mathbb M=\mathbb S^{N-1}\times_{\mathbb Z_2} \mathbb
S^1$, can be obtained as the coset $\mathbb M=SO(N,2)/\mathbb W$,
where $\mathbb W$ denotes the Weyl subgroup generated by $K_\mu,
M_{\mu\nu}$ and $D$ (i.e., a Poincaré subgroup $\mathbb P=SO(N-1,1)\circledS
\mathbb R^N$ augmented by dilations $\mathbb R^+$). The Weyl group
$\mathbb W$ is the stability subgroup (the little group in physical usage)
of $x^\mu=0$.

For $N=2$,  the group $SO(2,2)$  is isomorphic to the direct product
$SO(1,2)\times SO(1,2)$. It is well known that, in two dimensions, the
conformal group is infinite dimensional. Actually, the splitting
$SO(2,2)=SO(1,2)\times SO(1,2)$ has to do with the separation into
holomorphic and anti-holomorphic self-maps of the infinitesimal conformal
isometries of a complex domain, the generators of which,
\be
L_n=-z^{n+1}\frac{\partial}{\partial z},\, \bar L_n=-\bar z^{n+1}\frac{\partial}{\partial \bar z}, \, z=x^1+ix^0,  \bar z=x^1-ix^0,\,
n\in \mathbb Z,\ee
close into the Witt algebra $[L_m,L_n]=(m-n)L_{m+n}$ (idem for $\bar L$).
The conformal group in $N=2+1$ dimensions, $SO(3,2)$, is also the symmetry
group of the anti de Sitter space in $3+1$-dimensions,
AdS$_4=SO(3,2)/SO(3,1)$, a maximally symmetric Lorentzian manifold with
constant negative scalar curvature (i.e., the Lorentzian analogue of
4-dimensional hyperbolic space) which arises, for instance, as a vacuum
solution of Einstein's General Relativity field equations with a negative
(attractive) cosmological constant (corresponding to a negative vacuum
energy density and positive pressure).

We shall focus on the 15-parameter conformal group in $3+1$-dimensions, $SO(4,2)$, which turns out to be locally
isomorphic to the pseudo-unitary group
\be SU(2,2)=\left\{g\in {\rm Mat}_{4\times 4}(\mathbb C):  g^\dag \Gamma
g=\Gamma, \det(g)=1\right\}\label{su22gen} \ee
of complex special ${4\times 4}$ matrices $g$ leaving invariant  the
${4\times 4}$ hermitian form $\Gamma$ of signature $(++--)$. Here $g^\dag$
stands for adjoint (or conjugate/hermitian transpose) of $g$ (it is also
customary to denote it by $g^*$). Actually, the conformal Lie algebra
(\ref{conformalgebra}) can be also realized in terms of the Lie algebra
generators of the fundamental representation of $SU(2,2)$, given by the
following $4\times 4$ matrices
\be\ba{rcl} D&=&\frac{\gamma^5}{2},\;M^{\mu\nu}=\frac{\lb
\gamma^\mu,\gamma^\nu\rb}{4}=\frac{1}{4}\left(\ba{cc} \sigma^\mu\check{\sigma}^\nu-\sigma^\nu\check{\sigma}^\mu & 0\\
0&\check{\sigma}^\mu\sigma^\nu-\check{\sigma}^\nu\sigma^\mu\ea\right),\\
P^\mu&=&\gamma^\mu\frac{1+\gamma^5}{2}=\left(\ba{cc} 0& \sigma^\mu \\ 0
&0\ea\right),\;K^\mu=\gamma^\mu\frac{1-\gamma^5}{2}=\left(\ba{cc} 0& 0
\\ \check{\sigma}^\mu &0\ea\right)\ea\label{confalgamma}\ee
where
\be \gamma^\mu=\left(\ba{cc} 0& \sigma^\mu \\ \check{\sigma}^\mu
&0\ea\right),\;\; \gamma^5=i\gamma^0\gamma^1\gamma^2\gamma^3=\left(\ba{cc}
-\sigma^0& 0\\ 0& \sigma^0\ea\right),\nn\ee
denote the Dirac gamma matrices in the Weyl basis and
\be \sigma^0\equiv I=\left(\ba{cc} 1& 0
\\ 0 &1\ea\right),\;\sigma^1=\left(\ba{cc} 0& 1
\\ 1 &0\ea\right),\;\sigma^2=\left(\ba{cc} 0& -i
\\ i &0\ea\right),\;\sigma^3=\left(\ba{cc} 1& 0
\\ 0 &-1\ea\right),\label{Paulimat}\ee
are the Pauli matrices (we are writing $\check{\sigma}^\mu\equiv
\sigma_\mu$). Indeed, using standard properties of gamma and Pauli
matrices, one can easily check that the choice (\ref{confalgamma})
fulfils the commutation relations (\ref{conformalgebra}).

To be more precise, $SU(2,2)$ is the four-cover of $SO(4,2)$, much in the
same way as $SU(2)$ is the two-cover of $SO(3)$. This local isomorphism
between the conformal group $SO(N,2)$ and the pseudo-unitary group
$SU(M,M)$ only happens for $N=1$ and $N=4$ dimensions, where
\be
SO(1,2)=SU(1,1)/\mathbb Z_2,\; SO(4,2)=SU(2,2)/\mathbb Z_4.\ee
The $\lambda$-extension of the SMT given in Theorem \ref{GMSMT2}
(and Theorem \ref{MSMTN}) turns out to be closely related to the
group $SU(2,2)$ (and $SU(N,N)$ in general), providing a kind of
generating function for the unitary-representation functions of
this group (the discrete series, to be more precise). This formula
will be a useful mathematical tool for us, specially in proving
admissibility and tight-frame conditions and providing
reconstruction formulas. From this point of view, the conformal
group $SO(N,2)$ in $N=4$ dimensions is singled out from the
general $N$-dimensional case, at least in this article.

Before tackling the construction of conformal wavelets in 8-dimensional
complex Minkowski space in Section \ref{CWS}, we shall briefly remind the
simpler case of the CWT on the time axis $\mathbb R$ and its extension to
the Lobachevsky plane $\mbC_+$ and the open unit disk $\mathbb D_1$, which
are homogeneous spaces of  $SO(1,2)$.

\section{Wavelets for the Affine Group}
\label{afin}

Let us consider the affine or similitude group of translations and
dilations in one dimension,
\be G=SIM(1)=\mathbb R\rtimes \mathbb R^+=\{ g=(b,a)/\,b\in\mathbb{R},
a\in \mathbb R^+\}, \nn\ee
with group law ($g''=g'g$):
\begin{equation}
\begin{array}{rcl}
a'' &=& a'a \\
b'' &=& b' + a'b
\end{array}\nn
\end{equation}
This group will serve us as an introduction for studying the most
interesting case of the conformal group $G=SO(4,2)$ as a ``similitude'' group
of space-time, which will be considered in the next section.

The left-invariant Haar measure is:
\be d\mu(g)=\frac{1}{a^2}da\wedge db \nn\ee
The representation
\be [{\cU}_\lambda(a,b)\phi](x)=a^{-\lambda}\phi(\frac{x-b}{a})\equiv
\phi_{a,b}(x) \label{repreafin}\ee
of $G$ on $L^2(\mathbb R,dx)$ is unitary for
$\lambda=\frac{1}{2}+is$. In fact, every ${\cU}_\lambda$ is
unitarily equivalent to ${\cU}_{1/2}$ and one always works with
$\lambda=\frac{1}{2}$. This representation is reducible and splits
into two irreducible components: the positive $\omega>0$ and
negative $\omega<0$ frequency subespaces. Restricting oneself to
the subspace $\omega>0$, the admissibility condition (\ref{norm})
assumes the form
\be \int_0^\infty \frac{|\hat\psi(\omega)|^2}{\omega} d\omega<\infty\nn\ee
where $\hat\psi$ stands for the Fourier transform of $\psi$.
Given an admissible function $\psi\in L^2(\mathbb{R},dx)$, the machinery
of wavelet analysis proceeds in the usual way.

\subsection{Wavelets on the Lobachevsky plane $\mbC_+$}

An extension of the representation (\ref{repreafin}) of the affine
group, this time on the space $L^2_h(\mathbb C_+,d\tilde\nu_\lambda)$ of
square integrable holomorphic functions on the upper half complex
plane (or forward tube domain)
\be\mathbb T_1\equiv \mbC_+\equiv \{w=x+iy\in\mathbb{C}/
\Im(w)=y>0\},\label{upperhalfplane}\ee
is given by:
\be [\tilde{\cU}_\lambda(a,b)\phi](w)=a^{-\lambda}\phi(\frac{w-b}{a}). \label{reprehalfplane}\ee
This representation is unitary with respect to the scalar product:
\begin{equation}
\langle
\phi|\phi'\rangle=\int_{\mbC^+}\overline{\phi(w)}\phi'(w)d\tilde\nu_\lambda(w,\bar
w),\;\; d\tilde\nu_\lambda(w,\bar
w)=\frac{2\lambda-1}{4\pi}\Im(w)^{2(\lambda-1)}
|dw|,\label{scprodtubo1}
\end{equation}
for any $\phi,\phi'\in L^2_h(\mathbb C_+,d\tilde\nu_\lambda)$, where we use $|dw|$ as a shorthand for the Lebesgue measure $d\Re(w)\wedge
d\Im(w)$. Although all representations $\tilde{\cU}_\lambda, \lambda\geq 1$, are equivalent, they become inequivalent
when the affine group is immersed
inside the conformal group of the time axis $\mathbb R$,
$SO(1,2)\simeq SL(2,\mathbb{R})\simeq SU(1,1)$. Actually, this will be the case with the
conformal group $SO(4,2)$ in the next section. This immersion of $SIM(1)$
inside $SL(2,\mathbb{R})$ is apparent for the Iwasawa decomposition KAN (see,
for instance, \cite{Barut}) when parameterizing and element  $g\in SL(2,\mathbb{R})$ as:
\begin{equation}
g=\left(\begin{array}{cc} \cos\theta &-\sin\theta \\
\sin\theta&\cos\theta \end{array}\right)
 \left(\begin{array}{cc} \frac{1}{\sqrt{a}}    &0 \\0& \sqrt{a} \end{array}\right)
\left(\begin{array}{cc} 1 &b \\0&1 \end{array}\right)
   =
   \left(\begin{array}{cc}
 \frac{\cos \theta }
   {{\sqrt{a}}} & \frac{b\,\cos\theta}
    {{\sqrt{a}}} - {\sqrt{a}}\,\sin \theta  \cr
    \frac{\sin \theta }{{\sqrt{a}}} &
   {\sqrt{a}}\,\cos \theta  +
   \frac{b\,\sin \theta }{{\sqrt{a}}}
   \end{array}\right)\nn
 \end{equation}
where $\theta\in(-\pi,\pi]$ (see \cite{acha} for a unified
group-theoretical treatment of wavelets on $\mathbb R$ and the
circle $\mathbb S^1$ inside $SL(2,\mathbb{R})$).

\subsection{Wavelets on the open unit disk $\mathbb D_1$}

There is a one-to-one mapping between the Lobachevsky plane
$\mbC_+$ and the open unit disk (or Cartan domain)
\be \mathbb D_1=\{z\in\mbC, |z|<1\},\label{disk}\ee
given through the Cayley transformation:
\be z(w)=\frac{1+iw}{1-iw} \leftrightarrow w(z)=i\frac{1-z}{1+z}.\label{Cayley1}\ee
Note that the (Shilov) boundary $\mathbb S^1=\{z\in \mbC: |z|=1\}$ of $\mbD_1$ is stereographically  projected onto the boundary
$\mathbb R=\{w\in \mbC: \Im(w)=0\}$ of $\mbC_+$ by $w(e^{i\theta})=\tan(\theta/2)$.

We can establish an isometry between
$L^2_h(\mbC_+,d\tilde\nu_\lambda)$ and the space $L^2_h(\mathbb
D_1,d\nu_\lambda)$ of square integrable holomorphic functions on
the unit disk $\mathbb D_1$ with integration measure
\be d\nu_\lambda(z,\bar z)=\frac{2\lambda-1}{\pi}(1-z\bar
z)^{2(\lambda-1)}|dz|,\;\; \lambda\geq 1, \ee
where $\bar z$ denotes complex conjugate. This isometry is given by the correspondence
\be\begin{array}{cccc} \cS_\lambda: &
L^2_h(\mbD_1,d\nu_\lambda)&\longrightarrow&
L^2_h(\mbC_+,d\tilde\nu_\lambda)\\
 & \phi & \longmapsto & \cS_\lambda\phi\equiv
\tilde\phi,\end{array} \nn\ee
with
\be\tilde{\phi}(w)=2^{2\lambda}(1-iw)^{-2\lambda}\phi(z(w))\label{isodisc}\ee
and $z(w)$ given by (\ref{Cayley1}). In fact, taking into account that $(1-z\bar
z)=2^2\Im(w)|1-iw|^{-2}$ and the Jacobian determinant
$|dz|/|dw|=2^2|1-iw|^{-4}$, then
\bea
\la\phi|\phi'\ra_{L^2_h(\mbD_1)}&=&\frac{2\lambda-1}{\pi}\int_{\mbD_1}\overline{\phi(z)}\phi'(z)
(1-z\bar z)^{2(\lambda-1)}|dz|\nn\\
&=&
\frac{2\lambda-1}{4\pi}\int_{\mbC_+}\overline{2^{2\lambda}(1-iw)^{-2\lambda}\phi(z(w))}
2^{2\lambda}(1-iw)^{-2\lambda}\phi'(z(w))\Im(w)^{2(\lambda-1)}|dw|\nn\\
&=&\int_{\mbC_+}\overline{\tilde{\phi}(w)}\tilde{\phi}(w)
d\tilde\nu_\lambda(w,\bar w)=\la\tilde\phi|\tilde\phi'\ra_{L^2_h(\mbC_+)}.\nn\eea
The constant factor ${(2\lambda-1)}/{\pi}$ of $d\nu_\lambda(z,\bar
z)$ is chosen so that the set of functions
\be \varphi_n(z)\equiv \binom{2\lambda+n-1}{n}^{1/2} z^n, \;\;
n=0,1,2,\dots,\label{basedisco} \ee
constitutes an orthonormal basis of $L^2_h(\mathbb D_1,d\nu_\lambda)$, as can be easily checked by direct computation.
These basis functions verify the following closure relation:
\be \sum_{n=0}^\infty \overline{\varphi_n(z)}\varphi_n(z')=(1-\bar zz')^{-2\lambda},\label{kerneldisco}\ee
which is nothing other than the \textit{reproducing (Bergman)
kernel} of $L^2_h(\mbD_1,d\nu_\lambda)$ (see e.g. \cite{Gazeau}
for a general discussion on reproducing kernels). We shall provide
a four-dimensional analogue of (\ref{basedisco}) and
(\ref{kerneldisco}) in Eqs. (\ref{basisfunc}) and
(\ref{reprodkernel}), respectively, and prove the orthonormality
in Appendix \ref{orthoap}. The isometry $\cS_\lambda$ given by
(\ref{isodisc}) maps the orthonormal basis (\ref{basedisco}) of
$L^2_h(\mathbb D_1,d\nu_\lambda)$ onto the orthonormal basis
\be\tilde{\varphi}_n(w)=2^{2\lambda}(1-iw)^{-2\lambda}\varphi_n(z(w)),\;
n=0,1,2,\dots\label{baseloba}\ee
of $L^2_h(\mathbb C_+,d\tilde\nu)$, which verify the reproducing
kernel relation
\be \sum_{n=0}^\infty
\overline{\tilde\varphi_n(w)}\tilde\varphi_n(w')=(\frac{i}{2}(\bar
w-w'))^{-2\lambda}.\label{kerneloba}\ee
Let us denote by $\cV_\lambda\equiv
\cS^{-1}_\lambda\tilde\cU_\lambda\cS_\lambda$ the representation
of the affine group on $L^2_h(\mbD_1,d\nu_\lambda)$ induced from
(\ref{reprehalfplane}) through the isometry (\ref{isodisc}). More
explicitly:
\be
[\cV_\lambda(a,b)\phi](z)=[\cS^{-1}_\lambda\tilde\cU_\lambda(a,b)\tilde\phi](z)=
a^{-\lambda}\left(\frac{1-i\frac{w(z)-b}{a}}{1-iw(z)}\right)^{-2\lambda}\phi\left(z\left(\frac{w(z)-b}{a}\right)\right).\nn\ee
%
%
This representation is, by construction, unitary on $L^2_h(\mathbb D_1,d\nu_\lambda)$.

\section{\label{CWS}Wavelets for the Conformal Group $SO(4,2)$}

The four-dimensional analogue of the extension of the time axis
$\mathbb R$ to the time-energy half-plane $\mbC_+$ is the
extension of the Minkowski space $\mathbb R^4$ to the
(eight-dimensional) future tube domain $\mbC^4_+$ of the complex
Minkowski space $\mbC^4$ (see later on this Section). The
four-dimensional analogue of the one-to-one mapping between the
half-plane $\mbC_+$ and the disk $\mathbb D_1$ is now the Cayley
transform (\ref{Cayley2}) between $\mbC^4_+$ and the Cartan domain
$\mathbb D_4=U(2,2)/U(2)^2$, the Shilov boundary of which is the
compactified Minkowski space $U(2)$ (the four-dimensional analogue
of the boundary $U(1)=\mathbb S^1$ of the disk $\mathbb D_1$). Let
us see all this mappings and constructions in more detail.

\subsection{Wavelets on the forward tube domain $\mbC^4_+$}

The four-dimensional analogue of the upper-half complex plane
(\ref{upperhalfplane}) is the future/forward tube domain
\be \mathbb T_4=\mbC^4_+\equiv\left\{W=X+iY=w_\mu\sigma^\mu\in
{\rm Mat}_{2\times 2}(\mbC):\, Y>0\right\}\label{tube}\ee
of the complex Minkowski space $\mbC^4$, with $X=x_\mu\sigma^\mu$
and $Y=y_\mu\sigma^\mu$ hermitian matrices fulfilling the
positivity condition $Y>0\Leftrightarrow
y^0=\Im(w^0)>\|\vec{y}\|$.

The domain $\mbC^4_+$ is naturally homeomorphic to the quotient
$SU(2,2)/S(U(2)\times U(2))$ in the realization of the conformal
group in terms of $4\times 4$ complex (block) matrices $f$
fulfilling
\be G=SU(2,2)=\left\{f=\left(\ba{cc} R& iS
\\ -iT &Q\ea\right)\in {\rm Mat}_{4\times 4}(\mbC):  f^\dag \Gamma f=\Gamma,
\det(f)=1\right\},\label{su22} \ee
with
\be \Gamma=\gamma^0=\left(\ba{cc} 0& I
\\ I  & 0\ea\right)\nn\ee
the time component of the Dirac $4\times 4$ matrices $\gamma^\mu$ in the
Weyl basis ($I=\sigma^0$ is the $2\times 2$ identity matrix and
$f^\dag=f^*$ stands again for adjoint or conjugate/hermitian transpose of
$f$). In general, $\Gamma$ is a ${4\times 4}$ hermitian form of signature
$(++--)$. The inverse element of $f$ is then given by:
\be
f^{-1}=\gamma^0 f^\dag \gamma^0=\left(\ba{cc} Q^\dag& -iS^\dag
\\ iT^\dag &R^\dag\ea\right).\nn\ee
The particular identification of $\mbC^4_+$ with the coset
$SU(2,2)/S(U(2)^2)$ is given through:
\be
W=W(f)=(S+iR)(Q+iT)^{-1}=(Q+iT)^{-1}(S+iR).\label{tubecoord}\ee
The left translation $f'\to ff'$ of $G$ on itself induces a
natural left-action of $G$ on ${\mbC}^4_+$ given by:
\be W=W(f')\to W'=W(ff')=(RW+S)(TW+Q)^{-1}.\label{tubeaction}\ee
Let us make use of the standard identification $x_\mu\leftrightarrow
X=x_\mu\sigma^\mu$ between the Minkowski space $\mathbb R^{4}$ and the
space of $2\times 2$ hermitian matrices $X$, with $\sigma^\mu$ the Pauli
matrices (\ref{Paulimat}), and $x^2=x_\mu x^\mu=\det(X)$ the Minkowski
squared-norm. Setting $W=x_\mu\sigma^\mu$, the transformations
(\ref{confact}) can be formally recovered from (\ref{tubeaction}) as
follows:
\begin{itemize}
 \item[i)] Standard Lorentz transformations, $x'^\mu=\Lambda^{\mu}_\nu(\omega) x^\nu$, correspond to $T=S=0$ and
$R=Q^{-1\dag}\in SL(2,\mbC)$, where we are making use of the
homomorphism (spinor map) between $SO^+(3,1)$ and $SL(2,\mbC)$ and
writing $W'=RWR^\dag, R\in SL(2,\mbC)$ instead of
$x'^\mu=\Lambda^{\mu}_\nu x^\nu$. \item[ii)] Dilations correspond
to $T=S=0$ and $R=Q^{-1}=a^{1/2}I $ \item[iii)] Spacetime
translations correspond to $R=Q=I $ and $S=b_\mu\sigma^\mu, T=0$.
\item[iv)] Special conformal transformations correspond to $R=Q=I
$ and $T=c_\mu\sigma^\mu, S=0$ by noting that $\det(I
+TW)=1+2cx+c^2x^2$.
\end{itemize}

We shall give the next proposition without proof. Instead, we address the reader to
its counterpart (Proposition \ref{reprecartanprop})
in the next Subsection for an equivalent proof.

{\prop The representation of $G$ on square-integrable holomorphic
functions $\varphi(W)$ given by
\be [\tilde\cU_\lambda(f)\varphi](W)=\det(R^\dag-T^\dag W)^{-\lambda}
\varphi((Q^\dag W-S^\dag)(R^\dag-T^\dag
W)^{-1})\label{reprerest2}\ee
is unitary with respect to the integration measure
\be d\tilde\nu_\lambda(W,W^\dag)\equiv\frac{c_\lambda}{2^4}
\det(\frac{i}{2}(W^\dag-W))^{\lambda-4}|dW|=\frac{c_\lambda}{2^4}\Im(w)^{2(\lambda-4)}|dW|,\label{projintmeasure2}\ee
where $\lambda\in\mathbb N, \lambda>3$ (the ``{scale or conformal
dimension}'') is a parameter labelling non-equivalent
representations, $c_\lambda\equiv (\lambda-1)(\lambda-2)^2(\lambda-3)/\pi^4$
and we are using $|dW|=\bigwedge_{\mu=0}^3d\Re(w_\mu)d\Im(w_\mu)$
as a shorthand for the Lebesgue measure
 on $\mbC^4_+$.}

We identify the factor $\cM(f,W)^{1/2}=\det(R^\dag-T^\dag
W)^{-\lambda}$ in (\ref{reprerest2}) as a multiplier or
Radon-Nikodym derivative, (remember the general definition in
(\ref{multiplier})). It generalizes the factor $a^{-\lambda}$ in
(\ref{reprehalfplane}) by extending (global) standard dilations
$R=Q^{-1}=a^{1/2}I , T=S=0$ to (local/point-dependent)
``generalized dilations'' with $T=c_\mu\sigma^\mu$. The
representation (\ref{reprerest2}) is a special (spin-less or
scalar) case of the discrete series representations of $SU(2,2)$,
which are characterized by $\lambda$ and two spin labels $s_1$ and
$s_2$. Decomposing the discrete series representations of
$SU(2,2)$ into irreducible representations of the inhomogeneous
Lorentz group leads to a continuous (Poincaré) mass spectrum
\cite{Ruhl}.

\subsection{Wavelets on the Cartan domain $\mathbb D_4$\label{WCDsec}}

Instead of working in the forward tube domain $\mbC^4_+$, we shall
choose for convenience a different eight-dimensional space
$\mathbb D_4$ generalizing the (two-dimensional) open unit disk
$\mathbb D_1$ in (\ref{disk}), where we shall take advantage of
the full power of the $\lambda$-extension of the SMT given by the
formula (\ref{EMSMT}). Both spaces, $\mbC^4_+$ and $\mathbb D_4$,
are related by a Cayley-type transformation, which induces an
isomorphism between the corresponding Hilbert spaces of
square-integrable holomorphic functions on both manifolds (see
later on Section \ref{isometrysec}).

\subsubsection{Cayley transform and $\mathbb D_4$ as a coset of $SU(2,2)$}

The four-dimensional analogue of the map (\ref{Cayley1}) form the
Lobachevsky plane $\mbC_+$ onto the unit disk $\mathbb D_1$ is now
the Cayley transformation (and its inverse):
\be\ba{lll} W &\to& Z(W)=(I-iW)^{-1}(I+iW)=(I+iW)(I-iW)^{-1},\\
Z&\to&
W(Z)=i(I-Z)(I+Z)^{-1}=i(I+Z)^{-1}(I-Z),\ea\label{Cayley2}\ee
that maps (one-to-one) the forward  tube domain $\mbC^4_+$ onto
the Cartan complex domain defined by the positive-definite matrix
condition:
\be \mathbb D_4=\{Z\in {\rm Mat}_{2\times 2}(\mbC): I-ZZ^\dag>0\}.
\label{cartandomain}\ee
Note that the (Shilov) boundary
\[\check\mbD_4=U(2)=\{Z\in{\rm Mat}_{2\times 2}(\mbC): ZZ^\dag=Z^\dag Z=I\}=\mathbb S^3\times_{\mathbb Z_2}\mathbb S^1\]
of $\mbD_4$ is a compactification of the real Minkowski space
\[\mathbb M_4=\{W\in{\rm Mat}_{2\times 2}(\mbC): W^\dag=W\},\] i.e., the boundary of $\mbC^4_+$ (see e.g. \cite{Coquereaux}). The restriction
of the Cayley map $Z\to W(Z)$ to $Z\in U(2)$ is precisely the stereographic projection of $U(2)$ onto $\mathbb M_4$.

The Cartan domain $\mathbb D_4$ is naturally homeomorphic to the
quotient  $SU(2,2)/S(U(2)^2)$ in the new realization of:
\be G=SU(2,2)=\left\{g=\left(\ba{cc} A& B
\\ C &D\ea\right)\in {\rm Mat}_{4\times 4}(\mbC):  g^\dag \gamma^5 g=\gamma^5,
\det(g)=1\right\},\label{su22d} \ee
with
\be \gamma^5=\left(\ba{cc} -I & 0
\\ 0 & I  \ea\right)\nn\ee
the fifth Dirac $4\times 4$ gamma  matrix in the Weyl basis ($I=\sigma^0$ denotes again
the $2\times 2$ identity matrix). The inverse element of
$g$ is now:
\be
g^{-1}=\gamma^5 g^\dag \gamma^5=\left(\ba{cc} A^\dag& -C^\dag
\\ -B^\dag &D^\dag\ea\right).\nn\ee
Both realizations (\ref{su22}) and (\ref{su22d}) are
related by the map
\be f\to g=\left(\ba{cc} A& B
\\ C &D\ea\right)=\Upsilon^{-1}f\Upsilon=\frac{1}{2}\left(\ba{cc} R+iS-iT+Q &-R+iS+iT+Q\\
-R-iS-iT+Q & R-iS+iT+Q\ea\right),\label{upsilon3} \ee
with
\be
\Upsilon\equiv \frac{1}{\sqrt{2}}\left(\ba{cc} I & -I \\
I & I  \ea\right).\nn \ee
The particular identification of $\mathbb D_4$ with the coset
$SU(2,2)/S(U(2)^2)$ is given by [see later on Eq. (\ref{Iwasawa})
for more details]:
\be
Z(g)=BD^{-1},\;\; Z^\dag(g)=CA^{-1}.\label{domaincoord}
\ee
Actually, making explicit the matrix restrictions $g^\dag \gamma^5
g=\gamma^5$ in (\ref{su22d}):
\be g^{-1}g=I_{4\times 4}\Leftrightarrow \left\{\ba{r} D^\dag
D-B^\dag B=I\\ A^\dag A-C^\dag C=I\\ A^\dag B-C^\dag
D=0\ea\right.\label{mim}\ee
and
\be gg^{-1}=I_{4\times 4}\Leftrightarrow \left\{\ba{r} DD^\dag -CC^\dag
=I\\ AA^\dag -BB^\dag =I\\ AC^\dag- BD^\dag=0,\ea\right.\label{mmi}\ee
the positive matrix
condition in (\ref{cartandomain}) now reads
\be I-ZZ^\dag=I-A^{-1\dag}C^\dag
CA^{-1}=(AA^\dag)^{-1}>0,\label{positivityc}\ee
where we have used the second condition in (\ref{mim}).
Moreover, using the identification (\ref{domaincoord}) and the first condition in (\ref{mim}), we can see that
\be \det(ZZ^\dag)=\det(B^\dag B)\det(I+B^\dag
B)^{-1}<1,\label{dl1}\ee
This determinant restriction can also be proved as a direct consequence of
the positive-definite matrix condition $I-ZZ^{\dag}>0$. In fact, the
characteristic polynomial
\bea \det((1-\rho)I-ZZ^{\dag})&=& 1-\tr(\rho I+ZZ^{\dag})+\det(\rho
I+ZZ^{\dag})\nn\\ &=& \rho^2-(2-\tr(ZZ^{\dag}))\rho+\det(I-ZZ^{\dag})\nn
\eea
yields the eigenvalues
\be\rho_±=\frac{2-\tr(ZZ^{\dag})\pm\sqrt{\Delta}}{2},\;\;\Delta\equiv
(2-\tr(ZZ^{\dag}))^2 -4\det(I-ZZ^{\dag}). \label{eigenvalues}\ee
Since $I-ZZ^{\dag}$ is hermitian and positive definite, its eigenvalues
$\rho_±$ are real and positive. The condition $\rho_->0$ implies that:
\be 2-\tr(ZZ^{\dag})>0\Rightarrow \tr(ZZ^{\dag})<2\label{trless2},\ee
and the fact that $\Delta\geq 0$ gives:
\be 0\leq\Delta=\tr(ZZ^{\dag})^2-4\det(ZZ^{\dag})\Rightarrow
\det(ZZ^{\dag})\leq\frac{1}{4}\tr(ZZ^{\dag})^2<1,\label{dl1-2}\ee
where we have used (\ref{trless2}) in the last inequality. From
(\ref{trless2}), we can regard $\mathbb D_4$ as an open subset of the
eight-dimensional ball with radius $\sqrt{2}$. All those bounds for $Z\in
\mathbb D_4$ will be useful for proving convergence conditions later on
Section \ref{convergence}. See also the Appendix \ref{orthoap} for a
suitable parametrization of $Z$ when computing scalar products.

\subsubsection{Haar measure, unitary representation and reproducing kernel}

Any element $g\in G$  admits a Iwasawa
decomposition of the form
\be g=\left(\ba{cc} A& B
\\ C &D\ea\right)=\left(\ba{cc} \Delta_1& Z\Delta_2
\\ Z^\dag\Delta_1 &\Delta_2\ea\right)\left(\ba{cc} U_1& 0
\\ 0 &U_2\ea\right),\label{Iwasawa}\ee
with
\bea \Delta_1&=&(AA^\dag)^{1/2}=(I -ZZ^\dag)^{-1/2}, \;
U_1=\Delta_1^{-1}A,\nn\\
\Delta_2&=&(DD^\dag)^{1/2}=(I -Z^\dag Z)^{-1/2},
 U_2=\Delta_2^{-1}D. \nn\eea
This decomposition is adapted to the quotient $\mathbb D_4=G/H$ of
$G=SU(2,2)$ by the maximal compact subgroup $H=S(U(2)^2)$; that
is, $U_1,U_2\in U(2)$ with $\det(U_1U_2)=1$. In order to release
$U_{1,2}$ from the last determinant condition, we shall work from
now on with $G=U(2,2)$ and $H=U(2)^2$ instead. Likewise, a
parametrization of any $U\in U(2)$, adapted to the quotient
$\mathbb S^2=U(2)/U(1)^2$, is (the Hopf fibration)
\be U=\left(\ba{cc} a& b
\\ c &d\ea\right)=\left(\ba{cc} \delta & -z\delta
\\ \bar{z}\delta & \delta\ea\right)\left(\ba{cc} e^{i\beta_1}& 0
\\ 0 & e^{i\beta_2}\ea\right),\label{Iwasawa2}\ee
where $z=b/d\in \overline{\mbC}\simeq \mathbb S^2$ (the one-point
compactification of $\mbC$ by inverse stereographic projection),
$\delta=(1+z\bar{z})^{-1/2}$ and $e^{i\beta_1}=a/|a|,
e^{i\beta_2}=d/|d|$.

The left-invariant Haar measure [the exterior product of
left-invariant one-forms $g^{-1}dg$] of $G$ proves to be:
\bea
d\mu(g)&=&\left.d\mu(g)\right|_{G/H}\left.d\mu(g)\right|_{H},\label{haarmeasure}\\
\left.d\mu(g)\right|_{G/H}&=&
\det(I -ZZ^\dag)^{-4}|dZ|,\nn\\
\left.d\mu(g)\right|_{H}&=& dv(U_1)dv(U_2),\nn\eea
where we are denoting by $dv(U)$ the Haar measure on $U(2)$, which
[using (\ref{Iwasawa2})] can be in turn decomposed as:
\bea
dv(U)&\equiv& \left.dv(U)\right|_{U(2)/U(1)^2} \left.dv(U)\right|_{U(1)^2},\nn\\
\left.dv(U)\right|_{U(2)/U(1)^2}
 &=& \left.dv(U)\right|_{\mathbb{S}^2}\equiv ds(U)=(1+z\bar
z)^{-2}|dz|,\label{haarmeasures2}\\
\left.dv(U)\right|_{U(1)^2}&\equiv& d\beta_1 d\beta_2,\nn \eea
where $|dz|$ and $|dZ|$ denote the Lebesgue measures in  $\mbC$
and $\mbC^4$, respectively.

Let us consider the space of holomorphic functions $\phi(Z)$ on $\mathbb D_4$.

 {\prop \label{reprecartanprop} For any group element $g\in G$, the following (left-)action
\bea \phi_g(Z)&\equiv& [\cU_\lambda(g) \phi](Z)=\det(D^\dag-B^\dag
Z)^{-\lambda}\phi(Z'),\label{reprecartan}\\ Z'&=&g^{-1}Z=(A^\dag
Z-C^\dag)(D^\dag-B^\dag Z)^{-1} \nn\eea
defines a unitary irreducible square integrable representation of $G$ on
$L^2_h(\mathbb D_4,d\nu_\lambda)$ under the invariant scalar product
\bea\la \phi|\phi'\ra &=&\int_{\mathbb D_4}
\overline{\phi(Z)}\phi'(Z)d\nu_\lambda(Z,Z^\dag),\nn\\
d\nu_\lambda(Z,Z^\dag)&\equiv& c_\lambda
\det(I -ZZ^\dag)^{\lambda-4}|dZ| ,\label{scalarp}\eea
for any $\lambda\in\mathbb N,\, \lambda\geq 4$ (the ``{scale or
conformal dimension}''), where
$c_\lambda=\pi^{-4}(\lambda-1)(\lambda-2)^2(\lambda-3)$ is chosen
so that the unit function, $\phi(Z)=1, \,\forall Z\in\mathbb D_4$,
is normalized, i.e. $\la \phi|\phi \ra=1$. }

\ni \textbf{Proof:} One can easily check by elementary algebra
that $\cU_\lambda(g)\cU_\lambda(g')=\cU_\lambda(gg')$. In order to
prove unitarity, i.e. $\la \phi_g|\phi_g\ra=\la \phi|\phi\ra$ for
every $g\in G$, we shall make use of (\ref{mim}) and (\ref{mmi}).
In fact:
\be
\det(I-Z'Z'^\dag)=|\det(D^\dag-B^\dag Z)|^{-2}\det(I-ZZ^\dag), \nn\ee
and the Jacobian determinant
\be |dZ|=|dZ'| |\det(D^\dag-B^\dag Z)|^{8},\nn\ee
give the isometry relation $\|\phi_g\|^2=\|\phi\|^2$. Now taking
$g'=g^{-1}$ implies the unitarity of $\cU_\lambda$. For the computation of
$c_\lambda$ and other orthonormality properties see Appendix \ref{orthoap}
$\blacksquare$

In the next Section, we shall provide an isomorphism between
$L^2_h(\mbD_4,d\nu)$ and $L^2_h(\mbC^4_+,d\tilde\nu)$, where we
enjoy more physical intuition.

In order to prove admissibility conditions in Section \ref{admisec}, it will be convenient to give an
orthonormal basis of $L^2_h(\mathbb D_4,d\nu_\lambda)$.

{\prop The set of homogeneous polynomials of degree $2j+2m$:
\bea &\varphi_{q_1,q_2}^{j,m}(Z)\equiv\sqrt{\frac{2j+1}{\lambda-1}
\binom{m+\lambda-2}{\lambda-2}\binom{m+2j+\lambda-1}{\lambda-2}}\det(Z)^{m}\cD^{j}_{q_1,q_2}(Z),&
\label{basisfunc}\\  &m\in \mathbb N, \; j\in \mathbb N/2,
\;q_1,q_2=-j,-j+1,\dots,j-1,j,&\nn\eea
constitutes an orthonormal basis of $L^2_h(\mathbb D_4,d\nu_\lambda)$, that is:
\be\la\varphi_{q_1,q_2}^{j,m}|\varphi_{q'_1,q'_2}^{j',m'}\ra=
\delta_{j,j'}\delta_{m,m'}\delta_{q_1,q'_1}\delta_{q_2,q'_2}.\label{orthonormalityprop}\ee}

Note that the number of linearly independent polynomials
$\prod_{i,j=1}^2 z_{ij}^{n_{ij}}$ of fixed degree of homogeneity
$n=\sum_{i,j=1}^2n_{ij}$ is $(n+1)(n+2)(n+3)/6$, which coincides
with the number of linearly independent polynomials
(\ref{basisfunc}) with degree of homogeneity $n=2m+2j$. This proves
that the set of polynomials (\ref{basisfunc}) is a basis for
analytic functions $\phi\in L^2_h(\mathbb D_4,d\nu_\lambda)$. Moreover, this basis turns
out to be orthonormal. We address the interested reader to the Appendix \ref{orthoap} for a proof.
We prefer to omit it here in order to make
the presentation more dynamic.

Note also the close resemblance between the definition
(\ref{basisfunc}) and the left-hand side of the equality
(\ref{EMSMT}) in the $\lambda$-extended SMT \ref{GMSMT2}. In fact,
taking $tX=Z^\dag Z'$ in (\ref{EMSMT}) and using the properties
(\ref{Dmultprop}) and (\ref{transposprop}) of Wigner's
$\cD$-matrices, we can prove the following closure relation for
the basis functions (\ref{basisfunc}):
\be
\sum_{j\in\mbN/2}\sum^{\infty}_{m=0}\sum_{q,q'=-j}^j\overline{\varphi_{q',q}^{j,m}(Z)}\varphi_{q',q}^{j,m}(Z')
=\det(I-Z^\dag Z')^{-\lambda},\label{reprodkernel}\ee
which is nothing other than the \emph{reproducing (Bergman)
kernel} in $L^2_h(\mathbb D_4,d\nu_\lambda)$ (see e.g.
\cite{Gazeau} for a general discussion on reproducing kernels). Note that, although the scalar product
(\ref{scalarp}) is only valid for $\lambda\geq 4$, the expression (\ref{reprodkernel}) is formally valid for
$\lambda\geq 2$, since we are just using in it the requirements of the $\lambda$-extended SMT \ref{GMSMT2}. The
case $\lambda=2$ is related to the Szeg\"o kernel (see e.g. \cite{Coquereaux}).

\subsubsection{Admissibility condition, tight frame and reconstruction formula\label{admisec}}

{\thm \label{admissibthm} The representation (\ref{reprecartan}) is square integrable,
the constant unit function $\psi(Z)=\varphi^{0,0}_{0,0}(Z)=1, \,\forall Z\in\mathbb D_4$
being an admissible vector ({fiducial}
state or mother wavelet), i.e.:
\be
c_\psi=\int_G|\langle
\cU_\lambda(g)\psi|\psi\rangle|^2d\mu(g)<\infty\,\label{norm22}
\ee
and the set of {coherent states} (or {wavelets})
$F=\{\psi_g=\cU_\lambda(g)\psi, g\in G\}$ constituting a
continuous {tight frame} in $L^2_h(\mathbb D_4,d\nu_\lambda)$
satisfying the resolution of the identity:
\be {\cal A}=\int_G|\psi_g\ra\la\psi_g|d\mu(g)=c_\psi {\cal
I}\label{resolutop}.\ee
}
\ni\textbf{Proof:} Using the extended SMT \ref{GMSMT2} for
$tX=D^{-1}CZ$, we can expand
\bea \psi_g(Z)&=&\det(D^\dag-B^\dag Z)^{-\lambda}=\det(D^\dag)^{-\lambda}\det(I-(BD^{-1})^\dag Z)^{-\lambda}\nn\\
&=& \det(D^\dag)^{-\lambda}\sum_{j=0}^{\infty}\frac{2j+1}{\lambda-1}\sum^{\infty}_{n=0}
 \binom{n+\lambda-2}{\lambda-2}\binom{n+2j+\lambda -1}{\lambda-2}\label{expandcz}\\ & &\times
 \det((BD^{-1})^\dag Z)^{n}\sum_{q=-j}^j\cD^{j}_{qq}((BD^{-1})^\dag Z).\nn\eea
Now, taking into account that
$\det((BD^{-1})^\dag Z)^{n}=\det((BD^{-1})^\dag)^n\det(Z)^{n}$ and the property
(\ref{Dmultprop}) for
\be \cD^{j}_{qq}((BD^{-1})^\dag Z)=\sum_{q'=-j}^j
\cD^{j}_{qq'}((BD^{-1})^\dag)\cD^{j}_{q'q}(Z),\ee
we recognize the orthonormal basis functions (\ref{basisfunc}) in
the expansion (\ref{expandcz}), so that we can write the coherent
states (wavelets) as:
 \be
\psi_g(Z)=\sum_{j\in\mbN/2}\sum^{\infty}_{n=0}\sum_{q,q'=-j}^j\hat\psi_{q',q}^{j,n}(g)
\varphi_{q'q}^{j,n}(Z)\label{vexpansion}\ee
with  ``Fourier'' coefficients
 \bea \hat\psi_{q',q}^{j,n}(g)&=&\det(D^\dag)^{-\lambda}
 \sqrt{\frac{2j+1}{\lambda-1}\binom{n+\lambda-2}{\lambda-2}\binom{n+2j+\lambda
 -1}{\lambda-2}} \det((BD^{-1})^\dag)^{n}\nn\\ &&\times \sum_{q=-j}^j\cD^{j}_{qq'}((BD^{-1})^\dag)
= \overline{\det(D)^{-\lambda}
\varphi^{j,n}_{q',q}(BD^{-1})}.\label{Fourierpsi}\eea
Using the orthogonality properties (\ref{orthonormalityprop}) of the basis functions (\ref{basisfunc}), we can easily compute
\be
|\langle
\cU_\lambda(g)\psi|\psi\rangle|^2=|\langle
\psi_g|\varphi^{0,0}_{0,0}\rangle|^2=|\hat\psi^{0,0}_{0,0}(g)|^2=\det(DD^\dag)^{-\lambda}=\det(I -\tilde{Z}\tilde{Z}^\dag)^\lambda,
\ee
where we have defined $\tilde{Z}\equiv BD^{-1}$ and used the first condition in (\ref{mim}).
Using the Haar measure (\ref{haarmeasure}), the admissibility condition (\ref{norm22}) gives:
\be
c_\psi=\int_{G/H}\left.d\mu(g)\right|_{G/H}\det(I -\tilde{Z}\tilde{Z}^\dag)^\lambda\int_{H}dv(U_1)dv(U_2)=
c_\lambda^{-1}\left(\frac{(2\pi)^3}{2}\right)^2<\infty\,,\label{norm222}
\ee
where we have identified $\left.d\mu(g)\right|_{G/H}\det(I
-\tilde{Z}\tilde{Z}^\dag)^\lambda=
c_\lambda^{-1}d\nu_\lambda(\tilde{Z},\tilde{Z}^\dag)$ and taken into
account that  $\int_{\mathbb D_4}d\nu_\lambda(Z,Z^\dag)=1$ and
\be v(U(2))=\int_{U(2)}dv(U)=\int\frac{|dz|}{(1+z\bar
z)^{2}}d\beta_1 d\beta_2=\frac{(2\pi)^3}{2}\int_0^\infty
\frac{dx}{(1+x)^{2}}=\frac{(2\pi)^3}{2}, \ee
($2\pi$ times the area of the 3-sphere $\mathbb S^3= SU(2)$ of
unit radius).

Now it remains to prove that the resolution operator
(\ref{resolutop}) is a multiple of the identity ${\cal I}$ in
$L^2_h(\mathbb D_4,d\nu_\lambda)$. For this purpose, we shall
compute its matrix elements:
\bea
\la \varphi_{q_1,q_2}^{j,m}|{\cal A}|\varphi_{q'_1,q'_2}^{j',m'}\ra&=&\int_G\la
\varphi_{q_1,q_2}^{j,m}|\psi_g\ra\la\psi_g|\varphi_{q'_1,q'_2}^{j',m'}\ra d\mu(g)=
 \int_G \hat\psi_{q_1,q_2}^{j,m}(g)\overline{\hat\psi_{q'_1,q'_2}^{j',m'}(g)}d\mu(g)\nn\\
&=&
v(U(2))^2\int_{G/H}\left.d\mu(g)\right|_{G/H}\det(I -\tilde{Z}\tilde{Z}^\dag)^\lambda
\overline{\varphi_{q_1,q_2}^{j,m}(\tilde Z)} {\varphi_{q'_1,q'_2}^{j',m'}(\tilde Z)}\nn\\
&=&c_\psi \delta_{j,j'}\delta_{m,m'}\delta_{q_1,q'_1}\delta_{q_2,q'_2},
\eea
where we have used (\ref{Fourierpsi}), the orthogonality
properties (\ref{orthonormalityprop}) of the basis functions
(\ref{basisfunc}) and the fact that $G/H=\mathbb D_4$.
$\blacksquare$

The reconstruction formula (\ref{reconsform}) here adopts the following
form:
\be
\phi(Z)=\int_G\Phi_\psi(g)\psi_g(Z) d\mu(g),\label{reconsformD}\ee
with wavelet coefficients
\be
\Phi_\psi(g)=\frac{1}{c_\psi}\la\psi_g|\phi\ra=\frac{1}{c_\psi}\int_{\mathbb D_4}\det(D-Z^\dag B)^{-\lambda}\phi(Z)
d\nu_\lambda(Z,Z^\dag).\label{waveletcoef}\ee
Expanding $\phi$ in the basis (\ref{basisfunc})
\be
\phi(Z)=\sum_{j\in\mbN/2}\sum^{\infty}_{n=0}\sum_{q,q'=-j}^j\hat\phi_{q,q'}^{j,n}\varphi_{q,q'}^{j,n}(Z),\nn\ee
and using (\ref{vexpansion}) together with the orthogonality properties (\ref{orthonormalityprop}), we can write the wavelet coefficients
(\ref{waveletcoef}) in terms of the Fourier coefficients $\hat\phi_{q',q}^{j,n}$ as:
\be
\Phi_\psi(g)=\frac{1}{c_\psi}\sum_{j\in\mbN/2}\sum^{\infty}_{m=0}\sum_{q,q'=-j}^j\overline{\hat\psi_{q,q'}^{j,m}(g)}\hat\phi_{q,q'}^{j,m}=
\frac{1}{c_\psi}\sum_{j\in\mbN/2}\sum^{\infty}_{m=0}\sum_{q,q'=-j}^j\det(D)^{-\lambda} {\varphi^{j,m}_{q,q'}(BD^{-1})}\hat\phi_{q,q'}^{j,m}.\nn
\ee

\subsection{\label{isometrysec}Isomorphism between $L^2_h(\mbD_4,d\nu)$ and $L^2_h(\mbC^4_+,d\tilde\nu)$}

For completeness, we shall give an isometry between
$L^2_h(\mbD_4,d\nu)$ and $L^2_h(\mbC^4_+,d\tilde\nu)$ which allows
us to translate mathematical properties and constructions from one
space into the other.

{\prop\label{isometryDC} The correspondence
\be\begin{array}{cccc} \cS_\lambda: &
L^2_h(\mbD_4,d\nu_\lambda)&\longrightarrow&
L^2_h(\mbC^4_+,d\tilde\nu_\lambda)\\
 & \phi & \longmapsto & \cS_\lambda\phi\equiv
\tilde\phi,\end{array} \nn\ee
with
\be
\tilde\phi(W)=2^{2\lambda}\det(I-iW)^{-\lambda}\phi(Z(W))\label{isomap}\ee
and $Z(W)$ given by the Cayley transformation(\ref{Cayley2}), is
an isometry. Actually
\be
\la\phi|\phi'\ra_{L^2_h(\mbD_4,d\nu_\lambda)}=
\la\cS_\lambda\phi|\cS_\lambda\phi'\ra_{L^2_h(\mbC^4_+,d\tilde\nu_\lambda)}.\label{isoec}\ee
Moreover, $\cS_\lambda$ is an intertwiner (equivariant map) of the representations
(\ref{reprecartan}) and (\ref{reprerest2}), that is:
\be
\cU_\lambda=\cS_\lambda^{-1}\tilde\cU_\lambda\cS_\lambda.\label{intertwiner}\ee
}
\ni\textbf{Proof:} The left-hand side of the equality
(\ref{isoec}) is explicitly written as:
\be\la\phi|\phi'\ra_{L^2_h(\mbD_4,d\nu_\lambda)}=\int_{\mbD_4}\overline{\phi(Z)}\phi'(Z)c_\lambda
\det(I-ZZ^\dag)^{\lambda-4}|dZ|.\nn\ee
Taking into account that
\be\det(I-ZZ^\dag)=\det(2i(W^\dag-W))|\det(I-iW)|^{-2}\nn\ee
and the Jacobian determinant
\be |dZ|=2^{12}|\det(I-iW)|^{-8}|dW|,\nn\ee
then
\bea
d\nu_\lambda(Z,Z^\dag)&=&c_\lambda\det(I-ZZ^\dag)^{\lambda-4}|dZ|=\nn\\
&=&
2^{4\lambda-4}|\det(I-iW)|^{-2\lambda}{c_\lambda}\det(\frac{i}{2}(W^\dag-W))^{\lambda-4}
|dW|\nn\\ &=&
2^{4\lambda}|\det(I-iW)|^{-2\lambda}d\tilde\nu_\lambda(W,W^\dag),\nn\eea
which results in:
\be \int_{\mbD_4}\overline{\phi(Z)}\phi(Z)d\nu_\lambda(Z,Z^\dag)=
\int_{\mbC^4_+}\overline{\tilde\phi(W)}\tilde\phi(W)d\tilde\nu_\lambda(W,W^\dag),\nn\ee
thus proving (\ref{isoec}).

The intertwining relation (\ref{intertwiner}) can be explicitly
written as:
\be\ba{rll} [\cU_\lambda\phi](Z) &=&\det(D^\dag-B^\dag
Z)^{-\lambda}\phi\left((A^\dag Z-C^\dag)(D^\dag-B^\dag
Z)^{-1}\right)=\\
\left[\cS_\lambda^{-1}\tilde\cU_\lambda{\tilde\phi}\right](Z) &=&
\det(I-iW)^\lambda \det(R^\dag-T^\dag
W)^{-\lambda}\det(I-iW')^{-\lambda}\phi(Z(W')),
\ea\label{intertwiner2}\ee
where $W'=(Q^\dag W-S^\dag)(R^\dag-T^\dag W)^{-1}$. On the one
hand, we have that the argument of $\phi$ is:
\bea Z(W')&=&(I+iW')(I-iW')^{-1}\nn\\ &=& \left( (R^\dag-T^\dag
W)+i(Q^\dag W-S^\dag)\right)\left( (R^\dag-T^\dag W)-i(Q^\dag
W-S^\dag)\right)^{-1}\nn\\
&=&\left((R^\dag-iS^\dag )+i(Q^\dag
+iT^\dag)W\right)\left((R^\dag+iS^\dag )-i(Q^\dag
-iT^\dag)W\right)^{-1}.\nn\eea
Taking now into account the map (\ref{upsilon3})  we have:
\bea Z(W') &=&\left((A^\dag-C^\dag )+i(A^\dag
+C^\dag)W\right)\left((D^\dag-B^\dag )-i(D^\dag
+B^\dag)W\right)^{-1}\nn\\
&=&\left(A^\dag(I+iW)-C^\dag(I-iW)\right)\left(D^\dag(I-iW)-B^\dag(I+iW)\right)^{-1}\nn\\
&=&\left(A^\dag Z-C^\dag\right)\left(D^\dag-B^\dag
Z\right)^{-1},\nn\eea
as desired. On the other hand, we have that
\bea(I-iW')(R^\dag-T^\dag W)=(R^\dag-T^\dag W)-i(Q^\dag
W-S^\dag)=(R^\dag+iS^\dag )-i(Q^\dag -iT^\dag)W
 \nn\\  =(D^\dag-B^\dag )-i(D^\dag
 +B^\dag)W
 = D^\dag(I-iW)-B^\dag(I+iW)=(D^\dag-B^\dag
Z)(I-iW)\nn\eea
which implies
\be \det(I-iW)^\lambda \det(R^\dag-T^\dag
W)^{-\lambda}\det(I-iW')^{-\lambda}=\det(D^\dag-B^\dag
Z)^{-\lambda}\label{transvactub}\ee
That is, the equality of multipliers in (\ref{intertwiner2}). $\blacksquare$

As a direct consequence of Proposition \ref{isometryDC}, the set
of functions defined by
\be \tilde\varphi_{q_1,q_2}^{j,m}(W)\equiv
2^{2\lambda}\det(I-iW)^{-\lambda}\varphi_{q_1,q_2}^{j,m}(Z(W)),\label{basisfunc2}\ee
with $\varphi_{q_1,q_2}^{j,m}$ defined in (\ref{basisfunc}),
constitutes an orthonormal basis of
$L^2_h(\mbC^4_+,d\tilde\nu_\lambda)$ and the closure relation
\be
\sum_{j\in\mbN/2}\sum^{\infty}_{m=0}\sum_{q,q'=-j}^j\overline{\tilde\varphi_{q',q}^{j,m}(W)}
\tilde\varphi_{q',q}^{j,m}(W')
=\det(\frac{i}{2}(W^\dag-W'))^{-\lambda},\label{reprodkernelW}\ee
gives the reproducing (Bergman) kernel in $L^2_h(\mathbb
C^4_+,d\tilde\nu_\lambda)$.

The isometry (\ref{isomap}) also allows us to translate the results of Theorem \ref{admissibthm}
form $L^2_h(\mbD_4,d\nu_\lambda)$ into
$L^2_h(\mbC^4_+,d\tilde\nu_\lambda)$. Indeed, from (\ref{isomap}) we conclude that the function
$\tilde\psi\in L^2_h(\mbC^4_+,d\tilde\nu_\lambda)$ given by:
\be\tilde\psi(W)=2^{2\lambda}\det(I-iW)^{-\lambda}\label{admisswavtube}\ee
is admissible. The construction of a tight frame and a
reconstruction formula form this mother wavelet parallels
(\ref{resolutop}) and (\ref{reconsformD}), respectively.

\subsection{Symmetry properties of the proposed conformal wavelets}\label{symmetrysec}

When working with wavelets on the sphere
\cite{Holschneider-sphere,waveS2,nsphere} it is customary to take
\emph{axisymmetric} (or zonal) wavelets, that is, admissible vectors
$\psi$ which are invariant under rotations around the (namely) $z$-axis,
although more general implementations including directional spherical
wavelets are also possible (see e.g. \cite{directional}). Let us discuss
the symmetry properties of our proposed admissible wavelets
(\ref{admisswavtube}). Applying a general $SU(2,2)$-transformation
(\ref{reprerest2}) to (\ref{admisswavtube}) gives:
\be\ba{ll}
[\tilde\cU_\lambda(f)\tilde\psi](W)=2^{2\lambda}\det(R^\dag-T^\dag
W)^{-\lambda}\det(I-iW')^{-\lambda},\\ W'= (Q^\dag W-S^\dag)(R^\dag-T^\dag
W)^{-1}\ea\label{reprerest3}\ee
Using the identity (\ref{transvactub}) we have:
\be [\tilde\cU_\lambda(f)\tilde\psi](W)=\det(D^\dag-B^\dag
Z)^{-\lambda}\tilde\psi(W),\ee
which leaves invariant $\tilde\psi$ (up to a global phase) if:
\be B=0\Rightarrow C=0\Rightarrow S=-T,\; Q=R,\ee
where we have used (\ref{mim}), (\ref{mmi}) and  (\ref{upsilon3}). Thus, the
elements $f\in SU(2,2)$ of the form
\be f=\left(\ba{cc} R& iS
\\ iS &R\ea\right)\ee
leave invariant (\ref{admisswavtube}). The constraints $f^\dag \gamma^0 f=\gamma^0$ imply:
\be  S^\dag S+R^\dag R=I,\; S^\dag
R=R^\dag S.\ee
For $S=0$, $R$ is unitary. For $R=I$, $S$ is hermitian with
$S^2=0$. The last condition is satisfied for translations
$S=b_\mu\sigma^\mu$ along null (light-like) vectors $b^2=b_\mu
b^\mu=\det(S)=0$. This leaves us a 7-dimensional subgroup of
$SU(2,2)$, isomorphic to $S(U(2)\times U(2))$, as the isotropy
subgroup of the admissible vector (\ref{admisswavtube}). Any other
basis state (\ref{basisfunc2}) could be used as a fiducial state
to construct oriented wavelets.

In the next Figure we provide a visualization of this wavelet (modulus and argument) for the particular case of
$W=w\sigma^0$ (temporal part), $w\equiv x+iy$,  for which $\tilde\psi(W)=2^{2\lambda}(1-iw)^{-2\lambda}$ reduces to
$\tilde{\varphi}_0(w)$ in (\ref{baseloba}). We take $\lambda=1$ for simplicity.

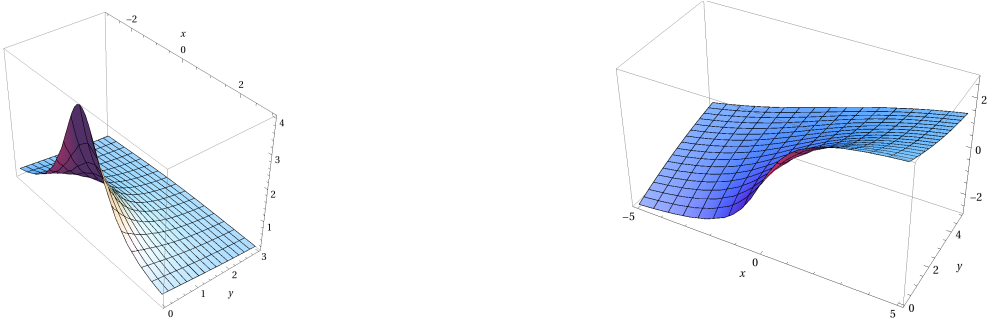
\begin{picture}(350,150)(0,0)
{\epsfxsize=5.0cm \put(0,0){\epsfbox{mother-abs.eps}}
\put(250,0){\epsfbox{mother-arg.eps}}}
\end{picture}

\subsection{The Euclidean limit}\label{euclideansec}

We have seen that the Shilov boundary of $\mathbb D_4$ is the compactified
Minkowski space $U(2)=\mathbb S^3\times_{\mathbb Z_2}\mathbb S^1$ (the
four-dimensional analogue of the boundary $U(1)=\mathbb S^1$ of the disk
$\mathbb D_1$). One expects the wavelet transform on $\mathbb S^N$ to
behave locally (at short scales or large values of the radius $\rho$) like
the usual (flat) wavelet transform on $\mathbb R^N$. Indeed, in
\cite{acha}, one of the authors and collaborators discussed the Euclidean
limit (infinite radius) for wavelets on $\mathbb S^1$. The procedure
parallels that of Ref. \cite{waveS2} for wavelets on $\mathbb S^2$. In
these references, the Euclidean limit is formulated as a contraction at
the level of group representations. Let us restrict ourselves, for the
sake of simplicity, to the conformal group $SO(1,2)$ in 1+0 (temporal)
dimensions. The realistic 1+3 dimensional case $SO(4,2)$, although
technically more complicated, follows similar guidelines and will be left
for future work.

Let us denote simply by $P=P_0$ and $K=K_0$ the temporal
components of $P_\mu$ and $K_\mu$ (the generators of spacetime translations and accelerations). The Lie algebra commutators
of $SO(1,2)$ are [remember the general $N$-dimensional case (\ref{conformalgebra})]:
\be [D,P]=-P,\; [D,K]=K,\; [K,P]=2D.\label{sl2rcom}\ee
A contraction ${\cal G}'$ of the Lie algebra  ${\cal G}=so(1,2)$ along sim$(1)$
(generated by $P$ and $D$) can be constructed through the
one-parameter family of invertible linear mappings
$\pi_\rho:\mathbb{R}^3\to\mathbb{R}^3, \rho\in [1,\infty)$ defined by:
\begin{equation}
\pi_\rho(D)=D,\;\pi_\rho(P)=P,\; \pi_\rho(K)=\rho^{-1}K,
\end{equation}
such that the Lie bracket of ${\cal G}'$ is:
\begin{equation}
[X,Y]'=\lim_{\rho\to\infty}\pi_\rho^{-1}[\pi_\rho X,\pi_\rho Y],
\end{equation}
with $[\cdot,\cdot]$ the Lie bracket (\ref{sl2rcom}) of
${\cal G}$. The resulting ${\cal G}'$ commutators  are:
 \begin{equation}
[D,P]'=-P,\; [D,K]'=K,\; [K,P]'=0.\label{sl2rcomp}
\end{equation}
The contraction process is lifted to the corresponding Lie groups $G'= \mathbb{R}^2\ltimes \mathbb{R}^+$
and $G=SO(1,2)$ by considering the exponential mapping $e^{\pi_\rho}$. The idea is that the representation of $G$ contract
to the usual wavelet representation of the affine group $SIM(1)$ in the
following sense:

\begin{defn}
Let $G'$ be a contraction of $G$, defined by the contraction map
$\Pi_\rho:G'\to G$, and let ${\cal U}'$ be a representation of $G'$ in a Hilbert space ${\cal H}'$.
Let $\{{\cal U}_\rho\}, \rho\in [1,\infty)$ be a one-parameter family of
representations of $G$ on a Hilbert space ${\cal H}_\rho$, and $\iota_\rho:{\cal
H}_\rho\to {\cal D}_\rho$ a linear injective map from ${\cal H}_\rho$ onto a dense
subspace ${\cal D}_\rho\subset {\cal H}'$. Then we shall say that ${\cal U}'$ is a
contraction of the family
$\{{\cal U}_\rho\}$ if there exists a dense subspace ${\cal D}'\subset{\cal H}'$  such that,
for all $\phi\in{\cal D}'$ and $g'\in G'$, one has:
\begin{itemize}
\item For every $\rho$ large enough, $\phi\in{\cal D}_\rho$ and
${\cal U}_\rho(\Pi_\rho(g'))\iota_\rho^{-1}\phi\in \iota_\rho^{-1}{\cal D}_\rho$ .
\item $\lim_{\rho\to\infty}||\iota_\rho{\cal U}_\rho(\Pi_\rho(g'))\iota_\rho^{-1}\phi-{\cal U}'(g')\phi||_{{\cal H}'}=0,\;\;\forall
g'\in G'$.
\end{itemize}
\end{defn}
More precisely, one can prove that:
\begin{thm}
The representation $[{\cal U}'(b,a)\phi](x)=\frac{1}{\sqrt{a}}\phi(\frac{x-b}{a})$ of
the affine group $SIM(1)$ is a contraction of the one-parameter family
${\cal U}_\rho$ of representations of $SO(1,2)$
on ${\cal H}_\rho$ as
$\rho\to\infty$. That is:
\begin{equation}
\lim_{\rho\to\infty}||\iota_\rho{\cal U}_\rho(\Pi_\rho^{\sigma'}(b,a))\iota_\rho^{-1}\phi-{\cal U}'(b,a)\phi||_{{\cal
H}'}=0,\;\;\forall (b,a)\in \mathbb{R}\ltimes \mathbb{R}^+,
\end{equation}
where $\Pi_\rho^{\sigma'}: SIM(1)\to SO(1,2)/\mathbb R$ is the restricted contraction map, with
$\sigma':G'/\mathbb R\to G'$ a given section.
\end{thm}
This construction can be straightforwardly extended to $G=SO(4,2)$, the contraction $G'$ being the
so-called $G_{15}$ group of Ref. \cite{conformecontract}. A thorough discussion of the Euclidean limit of the conformal
wavelets constructed in this paper falls beyond the scope of this article and will be left for future work \cite{euclidean4}.
Here we just wanted to give a flavor of it.

\section{\label{convergence} Convergence Remarks}

Schwinger's Theorem \ref{MSMT2} and its extension \ref{GMSMT2}
have been stated in the sense of generating functions in terms of
formal power series in some indeterminates. From this point of
view, we have disregarded convergence issues. However, infinite
series expansions like for instance (\ref{taylorexp}) would
require in particular that $|t|^2|\det(X)|<1$. We shall prove that
such convergence requirements, together with additional
restrictions coming from the basic Theorem \ref{MSMT2}, are
automatically fulfilled inside the complex domain $\mathbb D_4$
for $tX=\tilde Z^\dag Z$, with $\tilde Z= BD^{-1}$  in the
expansion (\ref{expandcz}). Let us state these convergence
requisites.

{\prop A sufficient condition for the convergence of the expansions
(\ref{MSMTF}) and (\ref{EMSMT}) for $t=1$ is that:
\be |x_{11}|<1,\; |x_{22}|<1,\; |x_{12}x_{21}|<1,\;
|\det(X)|<1.\label{matrixconditions}\ee}

\ni \textbf{Proof:} Looking at the explicit expression of Wigner's $\cD$-matrices
(\ref{Wignerf})
\be\cD^{j}_{q,q}(X)=\sum_{k=\max(0,2q)}^{j+q}
\binom{j+q}{k}\binom{j-q}{k-2q}x_{11}^k
(x_{12}x_{21})^{j+q-k}x_{22}^{k-2q}\label{Wignerf-q}\ee
we conclude that it is enough to have: $|x_{11}|<1$, $|x_{22}|<1$ and
$|x_{12}x_{21}|<1$, for the convergence of (\ref{MSMTF}) for $t=1$,
because their exponents run up to infinity independent of each other.
Moreover, if we require convergence in the expansions (\ref{EMSMT}) and
(\ref{taylorexp}) for $t=1$, then $|\det(X)|<1$ is needed too.
$\blacksquare$

We shall see that $Z$ and $\tilde Z$  fulfil
(\ref{matrixconditions}), but before we shall prove that

{\prop For any matrix $Z\in \mathbb D_4$ we have that the squared norm of
their rows is lesser than 1, that is:
\be |z_{11}|^2+|z_{12}|^2<1, \;
|z_{21}|^2+|z_{22}|^2<1.\label{ConvergenceConditions}\ee
}

\ni \textbf{Proof:} The positivity condition (\ref{positivityc}) says that
\be \det(I-ZZ^{\dag})>0 \Leftrightarrow
|z_{11}{\bar z_{21}}+z_{12}{\bar z_{22}}|^2<
(1-|z_{11}|^2-|z_{12}|^2)(1-|z_{21}|^2-|z_{22}|^2)\label{ConvergenceConditions2}.\ee
Hence, the last two factors must be either positive or negative. Supposing
that both factors were negative would contradict $\tr(ZZ^\dag)<2$ in
(\ref{trless2}). Therefore, we conclude that both factors are
positive.$\blacksquare$

Let us remind that, since $Z$ and $\tilde Z$ belong to $\mathbb D_4$, they must satisfy
$|\det(Z)|<1$ and $|\det(\tilde Z)|<1$, as we saw in (\ref{dl1}) and (\ref{dl1-2}).  Now we are in
condition to prove that:

{\prop The matrix $X=\tilde Z^\dag Z$ verifies the convergence conditions
(\ref{matrixconditions}) for every $Z,\tilde Z\in \mathbb D_4$ and,
therefore, the expansion (\ref{expandcz}) is well defined for
$\tilde Z=BD^{-1}$.}

 \ni \textbf{Proof:} The conditions
(\ref{ConvergenceConditions}) imply in particular that
$|z_{11}|<1,|z_{12}|<1,|z_{21}|<1,|z_{22}|<1$. Using this fact,
the triangle inequality and taking into account that
$Z,\tilde Z\in\mathbb D$ verify (\ref{ConvergenceConditions}) and the
determinant restriction (\ref{dl1-2}), we arrive to:
\bea |x_{11}|&=&|\tilde{z}_{11}{z_{11}}+\tilde{z}_{12}{z_{21}}|\leq
|\tilde{z}_{11}{z_{11}}| +|\tilde{z}_{12}{z_{21}}|<|\tilde{z}_{11}| +|\tilde{z}_{12}|
<|\tilde{z}_{11}|^2
+|\tilde{z}_{12}|^2<1,\nn\\
|x_{22}|&=&|\tilde{z}_{21}{z_{12}}+\tilde{z}_{22}{z_{22}}|\leq |\tilde{z}_{21}{z_{12}}|
+|\tilde{z}_{22}{z_{22}}|<|\tilde{z}_{21}| +|\tilde{z}_{22}| <|\tilde{z}_{21}|^2
+|\tilde{z}_{22}|^2<1,\nn\\
|x_{12}|&=&|\tilde{z}_{11}{z_{12}}+\tilde{z}_{12}{z_{22}}|\leq |\tilde{z}_{11}{z_{12}}|
+|\tilde{z}_{12}{z_{22}}|<|\tilde{z}_{11}| +|\tilde{z}_{12}| <|\tilde{z}_{11}|^2
+|\tilde{z}_{12}|^2<1,\nn\\
|x_{21}|&=&|\tilde{z}_{21}{z_{11}}+\tilde{z}_{22}{z_{21}}|\leq |\tilde{z}_{21}{z_{11}}|
+|\tilde{z}_{22}{z_{21}}|<|\tilde{z}_{21}| +|\tilde{z}_{22}| <|\tilde{z}_{21}|^2
+|\tilde{z}_{22}|^2<1,\nn\\
|\det(X)|&=&|\det(\tilde Z^\dag Z)|=|\det(\tilde
Z^\dag)||\det(Z)|={\det(\tilde Z^\dag\tilde
Z)}^{1/2}{\det(ZZ^\dag)}^{1/2}<1,\label{matrixcondproof}\eea
which proves the convergence conditions
(\ref{matrixconditions}).$\blacksquare$

\section{\label{conclu}Conclusions and Outlook}

We have constructed the CWT on the
Cartan domain $\mathbb D_4=
U(2,2)/U(2)^2$ of the conformal group $SO(4,2)=
SU(2,2)/\mathbb Z_4$ in 3+1 dimensions. The manifold $\mathbb D_4$
can be mapped one-to-one onto the future tube domain $\mbC^4_+$ of
the complex Minkowski space through a Cayley transformation, where
we enjoy more physical intuition. This construction paves the way
towards a new analysis tool of fields in complex Minkowski
space-time with continuum mass spectrum in terms of conformal
wavelets. It is traditional in Relativistic Particle Physics to
analyze fields or signals (for instance, elementary particles) in
Fourier (energy-momentum) space. However, like in music where
there are no infinitely lasting sounds, particles are created and
destroyed in nuclear reactions. A wavelet transform based on the
conformal group provides a way to analyze wave packets localized
in both: space and time. Important developments in this direction
have also been done in \cite{Kaiser1,Kaiser2,Kaiser3} for
electromagnetic (massless) signals.

In the way, we have stated and proved a $\lambda$-extension
(\ref{EMSMT}) of the Schwinger's formula (\ref{MSMTF}). This
extension turns out to be a useful mathematical tool for us,
specially as a generating function for the unitary-representation
functions of $SU(2,2)$, the derivation of the reproducing
(Bergman) kernel of $L^2_h(\mbD_4,d\nu_\lambda)$ and the proof of
admissibility and tight frame conditions. The generalization of
this theorem to matrices $X$ of size $N\geq 2$ follows similar
guidelines and the particular details are discussed in the
Appendix \ref{sizenap}, using the general $SU(N)$ solid harmonics
$\cD^p_{\alpha\alpha}(X)$ of Louck \cite{Louck2}. This result
could be of help in studying the discrete series
(infinite-dimensional) representations of the non-compact
pseudo-unitary groups $SU(N,N)$.

The next step should be the discretization problem. References
\cite{samplingsphere,samplinghyperboloid,Bogdanova} give us the
general guidelines to construct discrete (wavelet) frames on the
sphere and the hyperboloid and \cite{discretePoincare} on the
Poincaré group. The conformal group is much more involved, though
in principle the same scheme applies.

Looking for further potential applications of the conformal wavelets constructed in
this article, we think that they could be of use in analyzing
renormalizability problems in relativistic quantum field theory. When
describing space and time as a continuum, certain statistical and quantum
mechanical constructions are ill defined. In order to define them
properly, the continuum limit has to be taken carefully starting from a
discrete approach. There is a collection of techniques used to take a
continuum limit, usually referred as ``renormalization rules'', which
determine the relationship between parameters in the theory at large and
small scales. Renormalization rules fail to define a finite quantum theory
of Einstein's General Relativity, one of the main breakthroughs in
Theoretical Physics. The replacement of classical (commutative) space-time
by a quantum (non-commutative) space-time promises to restore finiteness
to quantum gravity at high energies and small (Planck) scales, where
geometry becomes also \emph{quantum} (non-commutative) \cite{Madore}.
Conformal wavelets could also be here of fundamental importance as an
analysis tool.

\section*{Acknowledgements}

Work partially supported by the Fundación Séneca (08814/PI/08),
Spanish MICINN (FIS2008-06078-C03-01) and Junta de Andaluc\'\i a
(FQM219). M.C. thanks the ``Universidad Politécnica de Cartagena''
and C.A.R.M.  for the award ``Intensificación de la Actividad
Investigadora''. ``Este trabajo es resultado de la ayuda concedida por la Fundación Séneca,
en el marco del PCTRM 2007-2010, con financiación del INFO y del FEDER de hasta un
80\%''. We all thank the anonymous referees for their comments which helped to improve the paper.


\appendix

\section{Extended MacMahon-Schwinger's Master Theorem for Matrices of Size $N \geq 2$\label{sizenap}}

We have shown the utility of the Theorem \ref{GMSMT2} in dealing
with unitary representations of $SU(2,2)$, in particular, in
proving the admissibility condition \ref{admissibthm}. We would
like to have a generalization of Theorem \ref{GMSMT2} for matrices
of arbitrary size $N$, since it would be a valuable tool as a
generating function for the unitary-representation functions of
$SU(N,N)$.

The first step is to generalize Wigner $\cD$-matrices. This
generalization has been done in the literature (see
\cite{Louck2} and references therein) by the so-called $SU(N)$ solid harmonics
$\cD^p_{\alpha\beta}(X)$ defined as:
\be \cD^p_{\alpha\beta}(X)\equiv\sqrt{\alpha!\beta!}\sum_{A\in
\mathbb M_{N}^p(\alpha,\beta)}
\frac{X^A}{A!}\label{SolidHarmonics}\ee
where the following space saving notations are employed: $A$ is a
$N\times N$ matrix in the nonnegative integers $a_{ij}$; $A!\equiv
\prod_{i,j=1}^N a_{ij}!$;  $X^A\equiv\prod_{i,j=1}^N
x_{ij}^{a_{ij}}$; $\alpha\equiv(\alpha_1,\alpha_2,...,\alpha_N)$
is a sequence of $N$ nonnegative integers that sum to $p$ (i.e., a
composition of $N$ into $p$ nonnegative parts), shortly
$\alpha\vdash p$; $\alpha!\equiv\prod_{i=1}^N \alpha_{i}!$;
$\mathbb M_{N}^p(\alpha,\beta)$ denotes the set of all matrices
$A$ such that the entries in row $i$ sum to $\alpha_i$ and those
in column $j$ sum to $\beta_j$, with $\alpha\vdash p$ and
$\beta\vdash p$. Hence, $\cD^p_{\alpha\beta}(X)$ are homogeneous
polynomials of degree $p$ in the indeterminates $x_{ij}$.

The particular identification with Wigner's $\tilde \cD$-matrices
for $X$ of size $N=2$ is given by
$\tilde\cD^{j}_{q_1,q_2}(X)=\cD^{p}_{\alpha\beta}(X)$ with $p=2j$,
$\alpha,\beta\vdash 2j$, $\alpha=(j+q_1,j-q_1)$, $\beta=(j+q_2,
j-q_2)$. Matrices $A\in\mathbb M_{2}^p(\alpha,\beta)$ can be then
indexed by an integer $k$
 \be
A^{(k)}\equiv\left(\ba{cc} k&j+q_1-k
\\ j+q_2-k &k-q_1-q_2\ea\right)\label{Ymatrix}\ee
with  $\max(0,q_1+q_2)\leq k \leq \min(j+q_1,j+q_2)$.

The multiplication property (\ref{Dmultprop}) and the
transpositional symmetry (\ref{transposprop}) for Wigner matrices
are still valid for $SU(N)$-solid harmonics as:
\be \sum_{\sigma\vdash p}
\cD^{p}_{\alpha\sigma}(X)\cD^{p}_{\sigma\beta}(Y)=\cD^{p}_{\alpha\beta}(XY)\label{Dmultpropap}\ee
 and
\be \cD^{p}_{\alpha\beta}(Y)=\cD^{p}_{\beta\alpha}(Y^T)\ee
(see \cite{Louck2} for a combinatorial proof).

Moreover, for
general $N\times N$ matrices $X$, the determinant $\det(I-X)$ can
be expanded in terms of sums of all principal $q$-th minors of $X$
as
\be \det(I-X)=\sum^N_{q=0}(-1)^{N+q}\sum_{\alpha\Vdash
q}\partial_x^{\alpha}\det(X),\nn
 \ee
where the $N$-dimensional multi-index
$\alpha\equiv(\alpha_1,\alpha_2,...,\alpha_N)$ is a partition of
$q$ with $\alpha_i\in\{0,1\}$, a fact that we now symbolize as
$\alpha\Vdash q$; $x\equiv(x_{11},x_{22},\dots,x_{NN})$ and
$\partial_x^\alpha=\prod_{i=1}^N\partial_{x_{ii}}^{\alpha_i}$. Let
us define the sum of all principal ($N-q$)-th minors of $X$ by
\be T_{q}(X)\equiv\sum_{\alpha\Vdash
N-q}\partial_x^\alpha\det(X).\nn\ee
They are homogeneous polynomials of degree $q=0,1\dots,N$ in the
indeterminates $x_{ij}$. For example: $T_0(X)= 1, T_1(X)={\rm
tr}(X),\dots,T_{N}(X)=\det(X)$.  Thus, $\det(I-X)$ can be written
in terms of these homogeneous polynomials as:
\be \det(I-X)=\sum^N_{q=0}(-1)^q T_{q}(X).\label{SeriesDet}\ee
Other possibility could be to use Waring's formulas \cite{Louck2}.

To arrive at the $\lambda$-extended MacMahon-Schwinger's Master
Theorem (MSMT) for $N\times N$ matrices, we shall now proceed step
by step from $\lambda=2$ to general $\lambda$. Before, let us
explicitly write down the generalization of the Theorem
\ref{MSMT2} to matrices of general size $N$

{\thm\label{MSMT2ap} {\rm (MSMT)} The identity
 \be
 \sum^{\infty}_{p=0}t^p\sum_{\alpha\vdash p}\cD^{p}_{\alpha\alpha}(X)={\det(I-tX)}^{-1}\label{MSMTFN}\ee
holds for any $N\times N$ matrix $X$.}

The action of the operator $D_1$ on both sides of the Basic
MacMahon-Schwinger's formula (\ref{MSMTFN}) now gives:
\be
 \sum^{\infty}_{p=0}(p+1)\sum_{\alpha\vdash
 p}\cD^{p}_{\alpha\alpha}(tX)=\frac{1-\sum^N_{q=2}(-1)^q(q-1)
 T_q(tX)}{\det(I-tX)^{2}}=\frac{1-\sum^N_{q=2}
 \widehat{T}_q(tX)}{\det(I-tX)^{2}},\label{trasD1-n}\ee
where we have defined $\widehat{T}_q(X)\equiv (-1)^q(q-1)T_q(X)$. We can
bring the numerator of the right-hand side of (\ref{trasD1-n}) back to the
left-hand side by using the expansion:
\be\frac{1}{1-\sum^N_{q=2}\widehat{T}_q(tX)}=\sum^{\infty}_{p=0}
(\sum^N_{q=2}\widehat{T}_q(tX))^{p}=\sum^{\infty}_{\gamma=0}\binom{\sum_{j=2}^N\gamma_j}{\gamma}
\widehat{T}(tX)^\gamma,\label{expan-n} \ee
where we have used the following shorthand for
\be \widehat{T}(X)^\gamma\equiv
\widehat{T}_2(X)^{\gamma_2}\widehat{T}_3(X)^{\gamma_3}
\dots\widehat{T}_N(X)^{\gamma_N},
\;\;\binom{\sum_{j=2}^N\gamma_j}{\gamma}\equiv
\frac{(\sum_{j=2}^N\gamma_j)!}{\gamma_2!\dots\gamma_N!}. \label{tgorro}\ee
Note that $\widehat{T}(X)^\gamma$ are homogeneous polynomials of
degree $\sum_{j=2}^N j\gamma_j$ in $x_{ij}$. Inserting the
expansion (\ref{expan-n}) in (\ref{trasD1-n}) we conclude that
\be
 \sum^{\infty}_{p=0}(p+1)\sum^{\infty}_{\gamma=0}\left\{\binom{\sum_{j=2}^N\gamma_j}{\gamma}\right\}
\widehat{T}(tX)^\gamma \sum_{\alpha\vdash
p}\cD^{p}_{\alpha\alpha}(tX)=\frac{1}{\det(I-tX)^{2}}.\label{auxlambda2}
 \ee
This is the generalization of (\ref{lambda2}) for general $N$. Let
us proceed by applying $D_2$ on both sides of the identity
(\ref{auxlambda2}):
\be
 \sum^{\infty}_{p=0}(p+1)\sum^{\infty}_{\gamma=0}\binom{\sum_{j=2}^N\gamma_j}{\gamma}
(p+2+\sum_{j=2}^N j\gamma_j)\widehat{T}(tX)^\gamma
\sum_{\alpha\vdash
p}\cD^{p}_{\alpha\alpha}(tX)=2\frac{1-\sum^N_{q=2}\widehat{T}_q(tX)}{\det(I-tX)^{3}}.\nn
 \ee
Using again (\ref{expan-n}), we have
 \bea
 \sum^{\infty}_{p=0}\frac{p+1}{2}\sum^{\infty}_{\gamma=0}\sum^{\infty}_{\gamma'=0}
\binom{\sum_{j=2}^N\gamma_j}{\gamma}\binom{\sum_{j=2}^N\gamma_j'}{\gamma'}
(p+2+\sum_{j=2}^N j\gamma_j)\widehat{T}(tX)^{\gamma+\gamma'}
\sum_{\alpha\vdash
p}\cD^{p}_{\alpha\alpha}(tX)\nn\\=\frac{1}{\det(I-tX)^{3}}\label{auxlambda3}
 \eea
Rearranging series as in (\ref{reseries}) and making the change of
$(N-1)$-dimensional multi-index: $\sigma\equiv\gamma+\gamma'$, we
 obtain
 \bea
 \sum^{\infty}_{p=0}\frac{p+1}{2}\sum^{\infty}_{\sigma=0}\left\{
 \sum^{\sigma}_{\gamma=0}\binom{\sum_{j=2}^N\gamma_j}{\gamma}\binom{\sum_{j=2}^N(\sigma_j-\gamma_j)}{\sigma-\gamma}
(p+2+\sum_{j=2}^N j\gamma_j)\right\}\nn\\
\times \widehat{T}(tX)^{\sigma}\sum_{\alpha\vdash
p}\cD^{p}_{\alpha\alpha}(tX) =\frac{1}{\det(I-tX)^3}.\label{auxlambda3b}
 \eea
Applying now $D_3$ on both sides of (\ref{auxlambda3}) results:
\bea
 \sum^{\infty}_{p=0}\frac{p+1}{2}\sum^{\infty}_{\gamma=0}\sum^{\infty}_{\gamma'=0}
\binom{\sum_{j=2}^N\gamma_j}{\gamma}\binom{\sum_{j=2}^N\gamma_j'}{\gamma'}
(p+2+\sum_{j=2}^N j\gamma_j)(p+3+\sum_{j=2}^N
j(\gamma_j+\gamma_j'))\nn\\
\times\widehat{T}(tX)^{\gamma+\gamma'} \sum_{\alpha\vdash
p}\cD^{p}_{\alpha\alpha}(tX)
=3\frac{1-\sum^N_{q=2}\widehat{T}_q(tX)}{\det(I-tX)^{4}}\nn
 \eea
and using again (\ref{expan-n}) we get:
\bea
 \sum^{\infty}_{p=0}\frac{p+1}{3!}\sum^{\infty}_{\gamma=0}\sum^{\infty}_{\gamma'=0}
 \sum^{\infty}_{\gamma''=0} \binom{\sum_{j=2}^N\gamma_j}{\gamma}
\binom{\sum_{j=2}^N\gamma_j'}{\gamma'}\binom{\sum_{j=2}^N\gamma_j''}{\gamma''}(p+2+\sum_{j=2}^N
j\gamma_j)\nn\\
\times(p+3+\sum_{j=2}^N j(\gamma_j+\gamma_j'))
\widehat{T}(tX)^{\gamma+\gamma'+\gamma''} \sum_{\alpha\vdash
p}\cD^{p}_{\alpha\alpha}(tX)=\frac{1}{\det(I-tX)^{4}}.\label{auxlambda4}
 \eea
Rearranging series and making the change
$\sigma\equiv\gamma+\gamma'+\gamma''$, we can recast the last
expression as:
\bea
 \sum^{\infty}_{p=0}\frac{p+1}{3}\sum^{\infty}_{\sigma=0}\left\{\frac{1}{2}
 \sum^{\sigma}_{\gamma=0}\sum^{\sigma-\gamma}_{\gamma'=0}\binom{\sum_{j=2}^N\gamma_j}{\gamma}
 \binom{\sum_{j=2}^N\gamma_j'}{\gamma'}\binom{\sum_{j=2}^N(\sigma_j-\gamma_j-\gamma_j')}{\sigma-\gamma-\gamma'}\right.\nn\\
\times\left.(p+2+\sum_{j=2}^Nj\gamma_j)(p+3+\sum_{j=2}^N
j(\gamma_j+\gamma_j'))\right\}\widehat{T}(tX)^{\sigma}\sum_{\alpha\vdash
p}\cD^{p}_{\alpha\alpha}(tX)=\frac{1}{\det(I-tX)^{4}}.\nn
 \eea
If we repeat the process $(\lambda-4)$ more times, then we arrive at
the following identity:

\bea
 \sum^{\infty}_{p=0}\frac{p+1}{(\lambda-1)!}\sum^{\infty}_{\gamma=0}
 \sum^{\infty}_{\gamma'=0}\dots
 \sum^{\infty}_{\gamma^{(\lambda-2)}=0}\binom{\sum_{j=2}^N\gamma_j^{(\lambda-2)}}{\gamma^{(\lambda-2)}}
 \prod_{k=0}^{\lambda-3}\binom{\sum_{j=2}^N\gamma_j^{(k)}}{\gamma^{(k)}}\nn\\ \times (p+k+2+\sum_{j=2}^N
j\sum_{i=0}^{k}\gamma_j^{(i)})\widehat{T}(tX)^{\gamma+\gamma'+\dots+\gamma^{(\lambda-2)}}
\sum_{\alpha\vdash
p}\cD^{p}_{\alpha\alpha}(tX)=\frac{1}{\det(I-tX)^{\lambda}}.\nn
 \eea
Making once more the change
   $\sigma=\gamma+\gamma'+\dots+\gamma^{(\lambda-2)}$, we can write:
\be
 \sum^{\infty}_{p=0}\frac{p+1}{\lambda-1}\sum^{\infty}_{\sigma=0}
 C_{p,\sigma}^{\lambda}\widehat{T}(tX)^{\sigma} \sum_{\alpha\vdash
p}\cD^{p}_{\alpha\alpha}(tX)=\frac{1}{\det(I-tX)^{\lambda}}.\nn
 \ee
where we have defined the following coefficients:
 \bea
C_{p,\sigma}^{\lambda}&\equiv&\frac{1}{(\lambda-2)!}\sum^{\sigma}_{\gamma=0}
\sum^{\sigma-\gamma}_{\gamma'=0}
\dots\sum^{\sigma-\gamma-\dots-\gamma^{(\lambda-4)}}_{\gamma^{(\lambda-3)}=0}
\binom{\sum_{j=2}^N(\sigma_j-\sum_{i=0}^{\lambda-3}\gamma_j^{(i)})
}{\sigma-\sum_{i=0}^{\lambda-3}\gamma^{(i)}}\nn\\
&&\times\prod_{k=0}^{\lambda-3}\binom{\sum_{j=2}^N\gamma_j^{(k)}}{\gamma^{(k)}}(p+k+2+\sum_{j=2}^N
j\sum_{i=0}^{k}\gamma_j^{(i)}).\label{GeneralCoefficient}\eea
In order to account for the particular coefficients
$C_{p,\sigma}^2$ and $C_{p,\sigma}^3$, given inside curly brackets
in (\ref{auxlambda2}) and (\ref{auxlambda3b}), we must understand
in (\ref{GeneralCoefficient}) that: 1) summations on
${\gamma^{(k)}}$ with $k<0$ are absent, 2) empty or nullary sums
are zero and 3) empty or nullary products are 1, as customary.

Summarizing, we can enunciate the following:
{\thm\label{MSMTN} {\rm ($\lambda$-extended MSMT)} For every
$\lambda\in \mathbb{N}, \lambda \geq 2$ and every $N\times N$
matrix $X$, the following identity holds:
\be
 \sum^{\infty}_{p=0}\frac{p+1}{\lambda-1}\sum^{\infty}_{\sigma=0}t^{p+\sum_{j=2}^Nj\sigma_j}
 C_{p,\sigma}^{\lambda}
\widehat{T}(X)^{\sigma} \sum_{\alpha\vdash
p}\cD^{p}_{\alpha\alpha}(X)=\det(I-tX)^{-\lambda},
\label{MSMTF-G}\ee
with $C_{p,\sigma}^{\lambda}$ given by (\ref{GeneralCoefficient})
and $\widehat{T}(X)^{\sigma}$ by (\ref{tgorro}).}

The expression (\ref{MSMTF-G}) generalizes (\ref{EMSMT}) for
matrices $X$ of arbitrary size $N$. In fact, for $N=2$, the
coefficient (\ref{GeneralCoefficient}) reduces to:
\be
 C_{p,\sigma_2}^{\lambda}=\binom{\lambda-2+\sigma_2}{\lambda-2}
 \binom{\lambda-1+p+\sigma_2}{\lambda-2},\label{coef2}
 \ee
which agrees with (\ref{EMSMT}). We have also been able to find
simplifications of $C_{p,\sigma}^{\lambda}$ in the following cases
(we take the binomials in the generalized sense
$\binom{n}{m}={n(n-1)\dots(n-m+1)}/{m!}$ to account for fractional
$n$):
\begin{enumerate}
 \item[i)] For $2\leq\lambda\leq5$, the coefficients (\ref{GeneralCoefficient}) are given by:
 \bea
 C_{p,\sigma}^{\lambda}&=&\binom{\sum_{j=2}^N\sigma_j}{\sigma}\binom{\lambda-2+
 \sum_{k=2}^N \sigma_k}{\lambda-2}
 \left\{\binom{\lambda-1+p+\frac{1}{2}
 \sum_{k=2}^N k\sigma_k}{\lambda-2}\right.\nn\\ && \left.+\binom{\lambda-2+p+\frac{1}{2}
 \sum_{k=2}^N k\sigma_k}{\lambda-4}\frac{1}{4!}\sum_{k=3}^N (k-2) k\sigma_k\right\}\nn
 \eea
\item[ii)] For $N=3$ and $\lambda\geq 2$, the coefficients (\ref{GeneralCoefficient}) can be
 obtained trough the expression:
 \bea
 C_{p,\sigma}^{\lambda}&=&\binom{\sigma_2+\sigma_3}{\sigma}
 \binom{\lambda-2+\sigma_2+\sigma_3}{\lambda-2}\nn\\ & &\times \sum_{i=1}^{\frac{\lambda-\xi}{2}}
 \binom{\lambda-i+p+\sigma_2+\frac{3}{2}\sigma_3}{\lambda-2i}\prod_{j=1}^{i-1}
 \frac{\sigma_3-2(j-1)}{8j}\label{coef3}\eea
where we have defined $\xi\equiv {\rm Odd}(\lambda)$, that is, $\xi=0$
when $\lambda$ is even and $\xi=1$ when odd. See that (\ref{coef3})
reduces to (\ref{coef2}) for $\sigma_3=0$.

\end{enumerate}

\section{Continuous Wavelet Transform on a Manifold: a Brief\label{brief}}

The usual CWT on the real line $\mathbb{R}$ is derived from the
natural unitary representation of the affine group $G=SIM(1)$ in
the space of finite energy signals $L^2(\mathbb{R}, dx)$ (see
Section \ref{afin} for a reminder). The same scheme applies to the
CWT on a general manifold $\mbX$, subject to the transitive
action, $x\to gx, g\in G, x\in \mbX$, of some group of
transformations $G$ which contains dilations and motions on
$\mbX$. If the measure $d\nu(x)$ in $\mbX$ is $G$-invariant (i.e.
$d\nu(gx)=d\nu(x)$), then the natural left action of $G$ on
$L^2(\mbX,d\nu)$ given by:
\be
[\cU(g)\phi](x)=\phi(g^{-1}x), \;\;g\in G, \phi\in L^2(\mbX,d\nu),\label{repre}
\ee
defines a unitary representation, that is:
\be \langle
\cU(g)\varphi|\cU(g)\phi\rangle=\langle\varphi|\phi\rangle\equiv\int_\mbX\overline{\varphi(x)}\phi(x)d\nu(x).\nn\ee
When $d\nu$ is not strictly invariant (i.e.
$d\nu(gx)=\cM(g,x)d\nu(x)$), we have to introduce a \emph{multiplier} (Radon-Nikodym derivative)
\be
[\cU(g)\phi](x)=\cM(g,x)^{1/2}\phi(g^{-1}x), \;\;g\in G, \phi\in
L^2(\mbX,d\nu), \label{multiplier}\ee
in order to keep unitarity. The fact that
$\cU(g)\cU(g')=\cU(gg')$ (i.e. $\cU$ is a representation of $G$)
implies cohomology conditions for multipliers, that is:
\begin{equation}
\cM(gg',x)=\cM(g,x)\cM(g',g^{-1}x).  \label{1-cocycle}
\end{equation}
Consider now the space $L^2(G,d\mu)$ of square-integrable complex
functions $\Psi$ on $G$, where $d\mu(g)=d\mu(g'g),\,\forall g'\in G$,
stands for the left-invariant Haar measure, which defines the scalar
product
\be \left(\Psi|\Phi\right)=\int_G\overline{\Psi(g)}\Phi(g)d\mu(g).
\nn\ee
A non-zero function $\psi\in L^2(\mbX,d\nu)$ is called \emph{admissible}
(or a \emph{fiducial} vector) if $\Psi(g)\equiv \langle
\cU(g)\psi|\psi\rangle\in L^2(G,d\mu)$, that is, if
\be
c_\psi=\int_G\overline{\Psi(g)}\Psi(g)d\mu(g)=\int_G|\langle
\cU(g)\psi|\psi\rangle|^2d\mu(g)<\infty. \label{norm}
\ee

A unitary representation for which admissible vector exists is
called \textit{square integrable}. For a square integrable
representation, besides Eq. (\ref{norm}) the following property
holds (see \cite{GMP}):
\be
\int_G|\langle
\cU(g)\psi|\phi\rangle|^2d\mu(g)<\infty\,,\forall \phi\in L^2(\mbX,d\nu) \,.\label{norm2}
\ee

Let us assume that the representation $\cU$ is \emph{irreducible}, and that
there exists a function $\psi$ admissible, then a system of coherent
states (CS) of $L^2(\mbX,d\nu)$  associated to (or indexed by) $G$ is defined
as the set of functions in the orbit of $\psi$ under $G$
\be \psi_g\equiv\cU(g)\psi, \;\; g\in G.\nn \ee

There are representations without admissible vectors, since the
integration with respect to some subgroup diverges. In this case, or even
for convenience when admissible vectors exist, we can restrict ourselves
to a suitable homogeneous space $Q=G/H$, for some closed subgroup $H$.
Then, the non-zero function $\psi$ is said to be admissible
mod$(H,\sigma)$ (with $\sigma:Q\to G$ a given section) and the
representation $\cU$ square integrable mod$(H,\sigma)$, if the condition
\be \int_Q|\langle \cU(\sigma(q))\psi|\phi\rangle|^2
d\check{\mu}(q)<\infty,\;\;\forall \phi\in
L^2(\mbX,d\nu)\label{qsquare} \ee
holds, where $d\check{\mu}$ is a measure on $Q$ ``projected'' from
the left-invariant measure $d\mu$ on the whole $G$ (see
\cite{medida} for more details on this projection procedure). Note
that this more general definition of square integrability includes
the previous one for the trivial subgroup $H=\{e\}$ and $\sigma$
the identity function. The notions of square integrability and admissibility mod$(H,\sigma)$ were
introduced in \cite{AnnPoinc} (see also \cite{Gazeau}).

The coherent states indexed by $Q$ are defined as
$\psi_{\sigma(q)}=U(\sigma(q))\psi, q\in Q$, and they form an overcomplete
set in $L^2(\mbX,d\nu)$.

The condition (\ref{qsquare}) could also be written as an
``expectation value"
\be 0<\int_Q |\langle U(\sigma(q))\psi|\phi\rangle|^2
d\check{\mu}(q)=\langle \phi|\cA_\sigma |\phi\rangle <\infty
,\;\;\forall \phi\in L^2(\mbX,d\nu),\label{pbiop}\ee
 where
$\cA_\sigma=\int_Q|\psi_{\sigma(q)}\rangle\langle
\psi_{\sigma(q)}|d\check{\mu}(q)$ is a positive, bounded,
invertible operator. If the operator $\cA_\sigma^{-1}$ is also
bounded, then the set $F_\sigma=\{|\psi_{\sigma(q)}\rangle, q\in
Q\}$ is called a \emph{frame}, and a \emph{tight frame} if
$\cA_\sigma$ is a positive multiple of the identity, $\cA_\sigma=c
{I}, c>0$.

To avoid domain problems in the following, let us assume that $\psi$
generates a frame (i.e., that $\cA_\sigma^{-1}$ is bounded). The
\emph{Coherent State map}  is defined as the linear map
\be\begin{array}{cccc} \cT_\psi: & L^2(\mbX,d\nu)&\longrightarrow&
L^2(Q,d\check\mu)\\
 & \phi & \longmapsto &  \cT_\psi\phi\equiv\Phi_\psi,\end{array}
\label{cwt}\ee
with $\Phi_\psi(q)=\frac{\langle
\psi_{\sigma(q)}|\phi\rangle}{\sqrt{c_\psi}}$. Its range
$L^2_\psi(Q,d\check\mu)\equiv \cT_\psi(L^2(\mbX,d\nu))$ is
complete with respect to the scalar product
$(\Phi|\Phi')_\psi\equiv\left(\Phi|\cT_\psi \cA_\sigma^{-1}
\cT_\psi^{-1}\Phi'\right)_Q$ and $\cT_\psi$ is unitary from
$L^2(\mbX,d\nu)$ onto $L^2_\psi(Q,d\check\mu)$. Thus, the inverse
map $\cT_\psi^{-1}$ yields the \emph{reconstruction formula}
\be
\phi=\cT_\psi^{-1}\Phi_\psi=\int_Q\Phi_\psi(q)\cA_\sigma^{-1}\psi_{\sigma(q)}
d\check\mu(q),\;\;\Phi_\psi\in
L^2_\psi(Q,d\check\mu),\label{reconsform}\ee
which expands $\phi$ in terms of coherent states (wavelets)
$\cA_\sigma^{-1}\psi_{\sigma(q)}$ with wavelet coefficients
$\Phi_\psi(q)=[\cT_\psi\phi](q)$. These formulas acquire a simpler form
when $\cA_\sigma$ is a multiple of the identity, as it is precisely the
case considered in this article.

\section{Orthonormality of Homogeneous Polynomials\label{orthoap}}

In order to prove the orthonormality relations (\ref{orthonormalityprop}), we shall adopt the following decomposition for
a matrix $Z\in\mathbb{D}_4$
\be Z=U_1\varXi U_2^\dag,\nn\ee
where $U_{1,2}\in U(2)/U(1)^2$ [as in (\ref{Iwasawa2}) with
$\beta_1=\beta_2=0$] and  $\varXi={\rm diag}(\xi_1,\xi_2\}, \xi_{1,2}\in\mathbb D_1$.  This parametrization ensures that
$Z\in\mathbb{D}_4$; in fact
\be I-ZZ^\dag=
U_1(I-\varXi{\varXi}^\dag)U_1^\dag>0\ee
since the eigenvalues are $1-|\xi_{1,2}|^2>0$.

Let us perform this change of variables in the invariant measure
(\ref{scalarp}) of $L^2_h(\mathbb D_4,d\nu_\lambda)$. On the one
hand, the Lebesgue measure on $\mbC^4$ can be written as:
\be |dZ|=J|d\xi_1||d\xi_2| ds(U_1)ds(U_2),\nn\ee
with $ds(U_{1,2})$ defined in (\ref{haarmeasures2}) and
$J=\frac{1}{2}(|\xi_1|^2-|\xi_2|^2)^2$ is the Jacobian determinant.
The Lebesgue measures $|d\xi_{1,2}|$ and $|dz_{1,2}|$
will be written in polar coordinates
$\xi_k=\rho_ke^{i\theta_k}$ and $z_k=r_k
e^{i\alpha_k}, k=1,2$. On the other hand, the weight factor in (\ref{scalarp}) adopts the form
\be\det(I-ZZ^\dag)^{\lambda-4}=
 ((1-\rho_1^2)(1-\rho_2^2))^{\lambda-4}\equiv \Omega(\rho),\nn\ee
so that the invariant measure of  $L^2_h(\mathbb D_4,d\nu_\lambda)$ reads:
\bea d\nu_\lambda(Z,Z^\dag)&=& c_\lambda
J(\rho)\Omega(\rho)|d\xi_1||d\xi_2| ds(U_1)ds(U_2)\\
&=&c_\lambda
J(\rho)\Omega(\rho)\rho_1 d\rho_1 d\theta_1 \rho_2 d\rho_2 d\theta_2 (1+r_1^2)^{-2}r_1 dr_1 d\alpha_1 (1+r_2^2)^{-2}r_2 dr_2 d\alpha_2,\nn\eea
with $0\leq\rho_{1,2}<1, 0\leq r_{1,2}<\infty, 0\leq\theta_{1,2}<2\pi, 0\leq\alpha_{1,2}<2\pi$. Let us call \[\cN_{j,m}\equiv\sqrt{\frac{2j+1}{\lambda-1}
\binom{m+\lambda-2}{\lambda-2}\binom{m+2j+\lambda-1}{\lambda-2}}\] the normalization constants of the
basis functions (\ref{basisfunc}). We want to evaluate:
\be\la\varphi_{q_1,q_2}^{j,m}|\varphi_{q'_1,q'_2}^{j',m'}\ra= \cN_{j,m}\cN_{j',m'} \int_{\mbD_4}d\nu_\lambda(Z,Z^\dag)
\overline{\det(Z)^{m}\cD^{j}_{q_1,q_2}(Z)}\det(Z)^{m'}\cD^{j'}_{q'_1,q'_2}(Z).
\ee
Using determinant properties, Wigner's $\cD$-matrix properties
(\ref{Dmultprop}) and (\ref{transposprop}), and the fact that
$\det(U_{1,2})=1$ and $\varXi$ is diagonal, the previous
expression  can be restated as:
\bea\frac{\la\varphi_{q_1,q_2}^{j,m},\varphi_{q'_1,q'_2}^{j',m'}\ra}{\cN_{j,m}\cN_{j',m'}}&=& \sum_{q=-j}^j\sum_{q'=-j'}^{j'}c_\lambda\int_{\mbD_1^2}
J\Omega|d\xi_1||d\xi_2|
\cD^{j}_{q,q}(\overline{\varXi})
\cD^{j'}_{q',q'}(\varXi)\det(\overline{\varXi})^m\det(\varXi)^{m'}\nn\\ & &\times\int_{\mbS^2}
ds(U_1)\cD^{j}_{q_1,q}(\overline{U_1}) \cD^{j'}_{q'_1,q'}(U_1)\int_{\mbS^2} ds(U_2)
\cD^{j}_{q_2,q}(U_2) \cD^{j'}_{q'_2,q'}(\overline{U_2})\label{overlap-1}
\eea
Let us start evaluating the first integral.  For the diagonal matrix $\varXi$ we have that $\cD^{j}_{q_1,q_2}(\varXi)=\delta_{q_1,q_2}\xi_1^{j+q_1}
\xi_2^{j-q_1}$, so that
\bea\cD^{j}_{q,q}(\overline{\varXi})
\cD^{j'}_{q',q'}(\varXi)\det(\overline{\varXi})^m\det(\varXi)^{m'}=\overline{\xi_1}^{j+q}
\overline{\xi_2}^{j-q}\xi_1^{j'+q'} \xi_2^{j'-q'}
\overline{\xi_1}^m\overline{\xi_2}^m\xi_1^{m'}\xi_2^{m'}=\nn\\
=\rho_1^{j+j'+q+q'+m+m'}\rho_2^{j+j'-q-q'+m+m'}e^{i(j'-j+q'-q+m'-m)\theta_1}e^{i(j'-j+q-q'+m'-m)\theta_2}.\eea
Integrating out angular variables gives the restrictions
\bea\int_0^{2\pi}\int_0^{2\pi}\cD^{j}_{q,q}(\overline{\varXi})
\cD^{j'}_{q',q'}(\varXi)\det(\overline{\varXi})^m\det(\varXi)^{m'}d\theta_1
d\theta_2\nn\\ =4\pi^2\delta_{q,q'}
\delta_{j+m,j'+m'}\rho_1^{2(j+q+m)}\rho_2^{2(j-q+m)}.\nn\eea
Integrating the radial part:
\bea 4\pi^2 c_\lambda\int_0^1\int_0^1
J(\rho)\Omega(\rho)\rho_1^{2(j+q+m)}\rho_2^{2(j-q+m)}
\rho_1d\rho_1\rho_2d\rho_2\nn\\ =
\frac{(j+m)^2+(j+m+2q^2+1)\lambda-5q^2-1}
{\pi^2(\lambda-1)\tbinom{j+m+q+\lambda-1}{\lambda-1}
\tbinom{j+m-q+\lambda-1}{\lambda-1}}\equiv {\cal R}_{j+m}^{q}\nn\eea
and putting all together in (\ref{overlap-1}) we have:
 \bea \frac{\la\varphi_{q_1,q_2}^{j,m},\varphi_{q'_1,q'_2}^{j',m'}\ra}{\cN_{j,m}\cN_{j',m'}}=
 \delta_{j+m,j'+m'}\sum_{q=-\min\{j,j'\}}^{\min\{j,j'\}}{\cal R}_{j+m}^{q}\nn\\ \times
\int_{\mbS^2} ds(U_1)\cD^{j}_{q_1,q}(\overline{U_1})
\cD^{j'}_{q'_1,q}(U_1)\int_{\mbS^2} ds(U_2) \cD^{j}_{q_2,q}(U_2)
\cD^{j'}_{q'_2,q}(\overline{U_2}) \label{overlap-2}\eea
The last two integrals are easily computable. Actually they are a
particular case of the orthogonality properties of Wigner's
$\cD$-matrices. More explicitly:
\be\int_{\mbS^2} ds(U)\cD^{j}_{q_1,q_2}(\overline{U})
\cD^{j'}_{q'_1,q_2}(U)=\int_0^\infty\int_0^{2\pi}\frac{rdr d\alpha}{(1+r^2)^2}
\cD^{j}_{q_1,q_2}(\overline{U}) \cD^{j'}_{q'_1,q_2}(U)
=\delta_{j,j'}\delta_{q_1,q'_1} \frac{\pi}{2j+1}.\nn\ee
Going back to (\ref{overlap-2}) it results:
\be\la\varphi_{q_1,q_2}^{j,m}|\varphi_{q'_1,q'_2}^{j',m'}\ra=
\delta_{j,j'}\delta_{m,m'}\delta_{q_1,q'_1}\delta_{q_2,q'_2}(\frac{\cN_{j,m}}{2j+1})^2
\sum_{q=-j}^j \pi^2 {\cal R}_{j+m}^{q}.\nn\ee
Finally, taking into account the combinatorial identity:
\be \sum_{q=-j}^j (\lambda-1)\pi^2 {\cal R}_{j+m}^{q}=\frac{2j+1}{
\binom{m+\lambda-2}{\lambda-2}\binom{m+2j+\lambda-1}{\lambda-2}}\nn
\ee
and the explicit expression of the normalization constants $\cN_{j,m}$, we arrive at
the orthonormality relations (\ref{orthonormalityprop}).

\end{document}